# Simultaneous Observation of Carrier-Specific Redistribution and Coherent Lattice Dynamics in 2H-$MoTe_2$ with Femtosecond Core-Level Spectroscopy

Andrew R. Attar[1,2,*], Hung-Tzu Chang[3,*], Alexander Britz[1,2], Xiang Zhang[4], Ming-Fu Lin[2], Aravind Krishnamoorthy[5], Thomas Linker[5], David Fritz[2], Daniel M. Neumark[3,6], Rajiv K. Kalia[5], Aiichiro Nakano[5], Pulickel Ajayan[4], Priya Vashishta[5], Uwe Bergmann[1,†], Stephen R. Leone[3,6,7†]

1. Stanford PULSE Institute, SLAC National Accelerator Laboratory, Menlo Park, CA 94025, USA.
2. Linac Coherent Light Source, SLAC National Accelerator Laboratory, Menlo Park, CA 94025, USA.
3. Department of Chemistry, University of California, Berkeley, CA 94720, USA.
4. Department of Materials Science and NanoEngineering, Rice University, Houston, TX 77005, USA.
5. Collaboratory for Advanced Computing and Simulations, University of Southern California, Los Angeles, CA 90089, USA.
6. Chemical Sciences Division, Lawrence Berkeley National Laboratory, Berkeley, CA 94720, USA.
7. Department of Physics, University of California, Berkeley, CA 94720, USA

*****These authors contributed equally to this work

[†] Correspondence and requests for materials should be addressed to U.B. (email: bergmann@slac.stanford.edu) or S.R.L. (email: srl@berkeley.edu)

## Abstract

We employ few-femtosecond extreme ultraviolet (XUV) transient absorption spectroscopy to reveal simultaneously the intra- and interband carrier relaxation and the light-induced structural dynamics in nanoscale thin films of layered 2H-$MoTe_2$ semiconductor. By interrogating the valence electronic structure *via* localized Te 4*d* (39-46 eV) and Mo 4*p* (35-38 eV) core levels, the relaxation of the photoexcited hole distribution is directly observed in real time. We obtain hole thermalization and cooling times of 15±5 fs and 380±90 fs, respectively, and an electron-hole recombination time of 1.5±0.1 ps. Furthermore, excitations of coherent out-of-plane $A_{1g}$ (5.1 THz) and in-plane $E_{1g}$ (3.7 THz) lattice vibrations are visualized through oscillations in the XUV absorption spectra. By comparison to Bethe-Salpeter equation simulations, the spectral changes are mapped to real-space excited-state displacements of the lattice along the dominant $A_{1g}$ coordinate. By directly and simultaneously probing the excited carrier distribution dynamics and accompanying femtosecond lattice displacement in 2H-$MoTe_2$ within a single experiment, our work provides a benchmark for understanding the interplay between electronic and structural dynamics in photoexcited nanomaterials.

MoTe$_2$ is a member of the emerging class of two-dimensional layered transition metal dichalcogenide (TMDC) materials with atomically thin layers separated by weakly bound van der Waals interactions.[1–3] While MoTe$_2$ is stable in two structural phases at room temperature, including the semiconducting hexagonal (2H) and semimetallic monoclinic (1T') phases, the more thermodynamically stable 2H-phase MoTe$_2$ has attracted significant interest due to its bandgap of 0.9 eV in bulk and 1.1 eV in the monolayer, similar to that of Si (1.1 eV).[4] Recent investigations have shown the potential for integrating both monolayer and multilayer TMDC semiconductors into devices, including transistors,[5,6] photonic logic gates,[7] photodetectors,[8] and nonvolatile memory cells.[9] While the typical device application involves stacking different subcomponent thin-film materials into heterostructures to extend functionality,[7,9,10] the overall performance of TMDC-based devices can be dominated by the individual subcomponent's carrier dynamics.[11] For devices involving multilayer thin-films of 2H-MoTe$_2$, the carrier transport properties within the MoTe$_2$ subcomponent can dictate their functionality.[7,12] In addition to the carrier dynamics, there is a growing interest in the light-induced structural responses of TMDC's including MoTe$_2$ and WTe$_2$, especially for applications in optically controlled phase transitions.[13–15]

Recent studies have therefore aimed at characterizing the carrier relaxation and associated structural dynamics of few-layer and bulk 2H-MoTe$_2$.[14,16–18] Although the understanding of carrier-specific dynamics of holes and electrons is critical for designing ambipolar semiconductor devices,[12,19,20] separating the different contributions of the charge carriers by traditional time-resolved optical/IR and THz spectroscopies is challenging due to overlapping spectral features. This is exemplified by the studies of 2H-MoTe$_2$ by Li *et al*[16] and Chi *et al*,[17] where THz and optical experiments were used to measure the carrier lifetime, but without carrier specificity and without sensitivity to hot carrier intraband dynamics.

As a possible solution to these challenges, recent investigations have shown the potential of using ultrafast core-level spectroscopy such as XUV transient absorption to disentangle the intraband hole and electron dynamics in semiconductors.[21–25] In XUV transient absorption experiments, an optical pump pulse excites carriers across the bandgap at time zero and, at a series of controlled time delays, the energy-dependent change in carrier population is probed by core-level transitions to the partially occupied valence band (VB) and conduction band (CB). In some cases, photoexcited holes and electrons can be separately and simultaneously attributed to changes in spectrally distinct transitions from atomic core levels to the transiently empty states in the VB (holes) and reduced absorption in the transiently filled CB (electrons). These phenomena are collectively referred to here as "state-filling." It is possible not only to distinguish the hole and electron distributions, but to also gain energy-dependent dynamics of the specific hot carriers within the VB and CB.[22,26] The element-specificity inherent to core-level spectroscopy offers the additional capability to distinguish the individual subcomponent materials in heterostructures.[27,28] Finally, the sensitivity of core-level spectroscopy to bonding, bond distances, and symmetry provides simultaneous structural information of photoexcited materials.[29–34]

In the present study, we apply sub-5 fs XUV transient absorption spectroscopy on a 2H-$MoTe_2$ semiconductor thin film (~50 nm/70 layers) to probe the dynamics of intraband carrier-specific relaxation, interband electron-hole recombination, and excited-state coherent lattice displacement. We obtain element-specific information by simultaneously measuring the spectral range around the Te $N_{4,5}$ absorption edge (39-46 eV) and the Mo $N_3$ edge (35-38 eV), as a function of delay time after optical excitation. The XUV absorption captures the same total density of states information at both edges. This enables us to analyze the hole distribution dynamics in the VB, distinguishing between thermalization of the holes by carrier-carrier scattering, subsequent cooling by carrier-phonon scattering, and the much slower electron-hole recombination process. We also report on the observation of light-induced coherent lattice vibrations in the out-of-plane $A_{1g}$ and the in-plane $E_{1g}$ modes of 2H-$MoTe_2$, which are captured *via* high-frequency oscillations (THz) in the XUV spectra. By comparison to *ab initio* Bethe-Salpeter equation simulations of the core-level absorption spectrum with different 2H-$MoTe_2$ lattice geometries, the excited-state displacement of the lattice that drives the coherent motion is extracted. Using this approach, we are able to capture the real-space lattice displacement simultaneously with the carrier-carrier thermalization of the holes, cooling (hole-phonon), and electron-hole recombination, leading to a cohesive picture of carrier and structural dynamics in a layered semiconductor nanomaterial.

## Results and Discussion

### Experimental Scheme and Static Absorption of 2H-$MoTe_2$.

The experimental apparatus and methodology of the XUV transient absorption spectroscopy measurements performed here have been previously described[21] and a detailed description is provided in the Methods section and the Supporting Information. The experiment is performed on a thin-film (50 nm thick, *i.e.* 70 layers) polycrystalline sample of 2H-phase $MoTe_2$, which is synthesized by chemical vapor deposition (CVD) onto a $Si_3N_4$ susbtrate of 30 nm thickness and 3x3 $mm^2$ lateral size. The sample is characterized by Raman spectroscopy to confirm the 2H-phase structure (see Supporting Information for sample preparation and detailed characterization).[3] In the XUV transient absorption experiments, this sample is photoexcited by a sub-5 fs, visible-to-near infrared (VIS-NIR) pump pulse spanning 1.2-2.2 eV with an average photon energy of 1.65 eV. The initial photoexcited carrier density is ~$1x10^{20}$ $cm^{-3}$ (*i.e.* ~$7x10^{12}$ $cm^{-2}$ in each layer, given the ~0.7 nm thickness per layer), which corresponds to a 0.1% excitation fraction of the total valence electrons (see Supporting Information for details). The photoexcited sample is then probed at a controlled time delay, $\tau$, by the arrival of a broadband (30-50 eV), sub-5 fs XUV pulse produced by high-harmonic generation. The transmitted XUV intensity through the sample is measured as a function of photon energy, I(E), by a spectrometer. The transient absorption signal, $\Delta OD (E, \tau) = OD_{pump\text{-}on}(E, \tau) - OD_{pump\text{-}off}(E)$, is determined by the difference between the XUV absorbance, or optical density (OD), of the photoexcited sample at time $\tau$ and the static XUV absorbance in the absence of pump excitation.

The broadband XUV pulse allows for the simultaneous absorption measurement covering the Mo $N_{2,3}$ edges (~35-38 eV, referred to as the "Mo window") and the Te $N_{4,5}$-edges (39-46 eV, referred to as the "Te window"). The transmission of the broadband XUV spectrum (Figure S1c) is referenced to the transmission of a blank $Si_3N_4$ substrate to obtain the static absorption spectrum of the thin-film 2H-$MoTe_2$ sample: *i.e.* absorbance $A = \log(I_{blank}/I_{sample})$, which is shown in Figure 1a. XUV transitions in the Mo and Te windows occur by promotion of Mo $4p_{3/2,1/2}$ and Te $4d_{5/2,3/2}$ core electrons to the CB, respectively. The onset of the transitions from each observed core level is labeled 1-3 in Figure 1a and the corresponding transition is shown schematically in Figure 1b. Note that the arrows in Figure 1b only show the *onset* of the core-level absorption edge. As the XUV energy is increased above each onset, carriers are promoted from the corresponding core level to higher valleys in the CB at different k-space positions throughout the Brillouin zone. In the Mo window, the onset of transitions from the Mo $4p_{1/2}$ core level is too weak to observe in the static spectrum and is omitted from the analysis. The onset of Mo $4p_{3/2}$→CB absorption is measured at 36.1 eV and labeled edge 1. In the Te window, the onset of the Te $4d_{5/2}$→CB (edge 2) is measured at 40.7 eV and the onset of the Te $4d_{3/2}$→CB (edge 3) is at 42.2 eV. The 1.5 eV separation of edges 2 and 3 matches the Te $4d_{5/2,3/2}$ spin-orbit splitting in $MoTe_2$ confirmed by XPS spectra (Supporting Information Figure S3). When the VIS-NIR pulse excites the material, additional transitions become accessible from the core levels to the VB and transitions to the CB can be reduced due to state-filling, as discussed below.

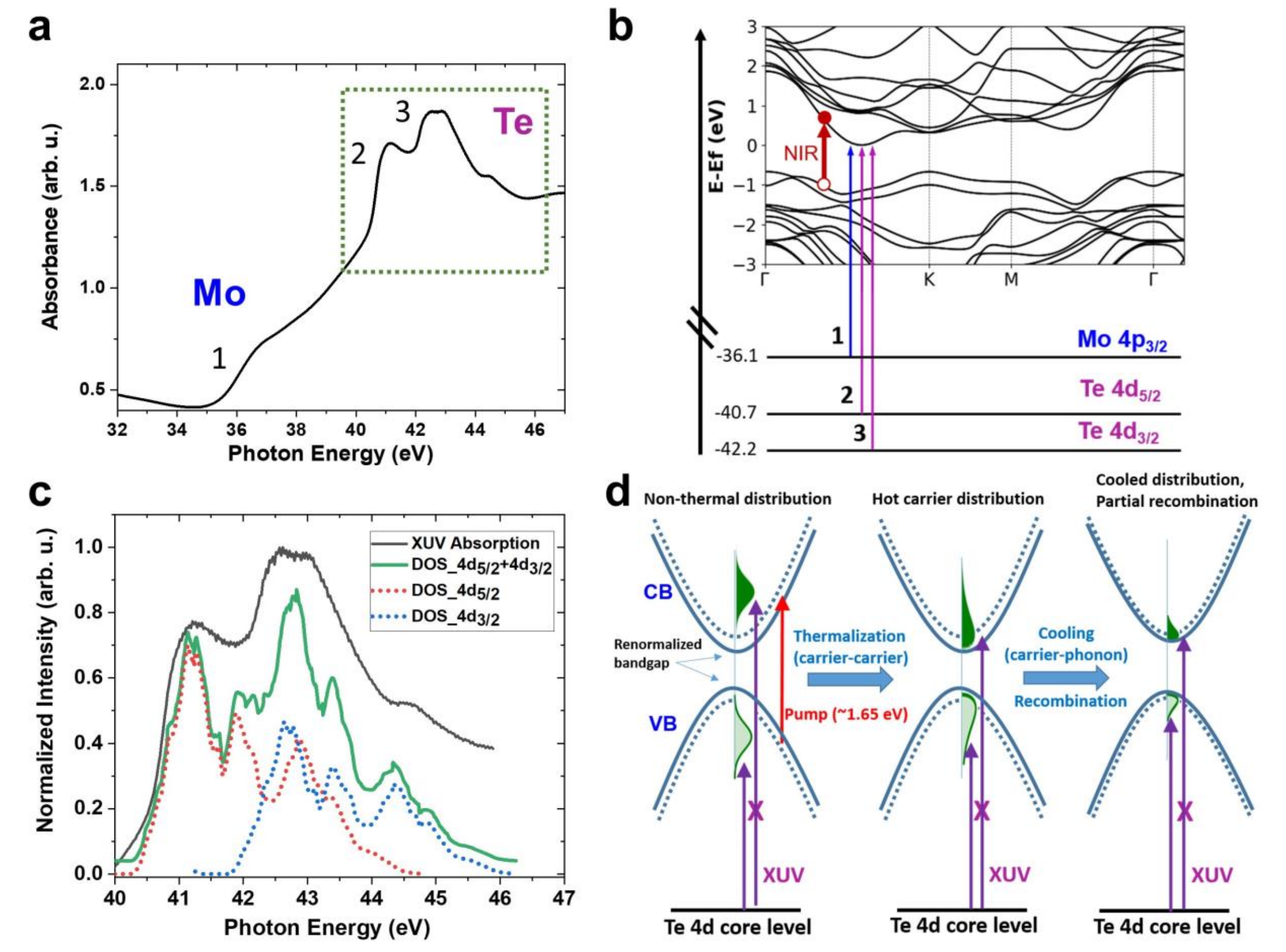

**Figure 1. (a)** Static XUV absorption spectrum of 2H-$MoTe_2$ with Mo $4p_{3/2}$→CB and Te $4d_{5/2,3/2}$→CB absorption edges labeled 1-3. The Mo $4p_{1/2}$→CB edge is too weak to observe and is omitted. **(b)** Band structure of 2H-$MoTe_2$ along the Γ–K–M–Γ path. The corresponding core-level transitions for the three absorption edges labeled in (a) are shown by vertical blue and purple arrows and labeled 1-3 accordingly. Note that the arrows only show the *onset* of the core-level absorption edge for the case of the material without NIR-VIS excitation. As the XUV energy is increased above each onset, carriers are promoted to higher energies in the CB at different positions throughout k-space. One representative VB→CB transition induced by the NIR-VIS pump pulse (of the range of transitions possible from the broad bandwidth pump) is shown as a red arrow. **(c)** An expanded plot of the normalized XUV absorption spectrum in the Te window (black line). The red-dotted and the blue-dotted lines are plots of the calculated normalized CB DOS relative to the Te $4d_{5/2}$ core level (labeled DOS_$4d_{5/2}$) and $4d_{3/2}$ core level (labeled DOS_$4d_{3/2}$), respectively. The difference in the relative amplitudes of DOS_$4d_{5/2}$ and DOS_$4d_{3/2}$ is described in the main text. The sum of DOS_$4d_{5/2}$ and DOS_$4d_{3/2}$ is plotted (green solid line). **(d)** Model of the XUV transient absorption probing scheme for 2H-$MoTe_2$. The VB and CB are shown as parabolic bands. The solid (dashed) parabolas represent the bands with (without) band-gap renormalization. The photoexcited holes and electrons are represented by the shaded light-green and dark-green areas, respectively. Representative XUV transitions from the Te 4d core levels are shown with and without an X to represent a decrease or increase in absorption relative to the static spectrum, respectively, due to state-filling.

Considering more closely the Te window (Figure 1c), the relatively sharp features allow for a detailed mapping of the CB density of states (DOS) from the Te 4*d* core levels. First, DOS calculations are performed, which are described and shown in detail in the Supporting Information, Figure S4. In Figure 1c, the calculated CB density of states (DOS) is plotted relative to the $4d_{5/2}$ core level (red dashed line, labeled DOS_$4d_{5/2}$) and to the $4d_{3/2}$ core level (blue dashed line, labeled DOS_$4d_{3/2}$). The relative amplitudes between DOS_$4d_{5/2}$ and DOS_$4d_{3/2}$ is set according to the expected degeneracies of the core-hole total angular momentum J states (J =5/2 *versus* J=3/2), which, to first approximation, predicts a 3:2 ratio for $4d_{5/2}$→CB *versus* $4d_{3/2}$→CB amplitudes. The resulting core-hole-mapped DOS (green solid line) is compared to the experimental XUV absorption spectrum (black solid line). The close agreement between the calculated CB DOS and the XUV absorption spectrum in Figure 1c demonstrates that the core hole of the Te $N_{4,5}$ absorption is well screened such that the core-level absorption spectrum can be regarded as a map of the CB unoccupied DOS in the valence shell.

Within this picture, the probing scheme for measuring the carrier dynamics following optical excitation can be understood by the model illustrated in Figure 1d. In the leftmost panel, above-bandgap photoexcitation produces a non-thermalized distribution of carriers corresponding to a convolution of the excitation spectrum and the VB/CB DOS. The broad bandwidth excitation in the present experiment produces carriers over a large range of momentum values in k-space. Moving from left to right in Figure 1d, the initial non-thermal carrier distribution is expected to undergo carrier-carrier thermalization, carrier-phonon cooling, and recombination in accordance with known carrier-carrier and carrier-phonon scattering processes in semiconductors.[35] These carrier dynamics lead to energy- and time-dependent changes in the unoccupied DOS within the VB and CB, as shown in the schematic. For example, the peak of the hot hole distribution is expected to shift upwards toward the VB maximum due to both thermalization and cooling, which can occur on timescales of 10's to 100's fs.[35–38] The corresponding changes to the XUV absorption (ΔOD) due to state-filling effects[21–23,26] are schematically illustrated by vertical arrows from the representative Te 4d core-level. In the following sections, the time-resolved ΔOD of the spectrally dispersed XUV probe is used to extract both the carrier distribution and population dynamics in 2H-$MoTe_2$, revealing the individual steps depicted in Figure 1d. Due to the sharper XUV absorption edge observed in the Te window as compared to the Mo window, we focus primarily on the Te N-edge to report on the carrier and structural dynamics in this work. However, as shown in the Supporting Information in Figure S6 and Figure S9, respectively, the Mo $4p_{3/2}$ core-level can also be used to capture the hole population dynamics and the coherent phonon dynamics.

**XUV Transient Absorption Following Broadband Photoexcitation.**

In Figure 2a, the change in the XUV absorbance, ΔOD(E, τ), following VIS-NIR excitation is plotted in a false-color map as a function of time delay and photon energy. Changes are observed in both the Mo and Te windows near 33-37 eV and 38-46 eV, respectively. The ground-state absorbance spectrum of 2H-$MoTe_2$ is shown here again in the upper panel of Figure 2a to emphasize where the major pump-induced absorption changes appear relative to the static spectrum. The dotted vertical lines in Figure 2a (upper panel) at 36.1 eV and 40.7 eV show the energies of the Mo $4p_{3/2}$→CB minimum and the Te $4d_{5/2}$→CB minimum, respectively. With a bulk bandgap of 0.9 eV in 2H-$MoTe_2$, the dash-dotted vertical lines at 35.2 eV and 39.8 eV correspond to the derived VB maximum energy relative to the Mo $4p_{3/2}$ and Te $4d_{5/2}$ core levels, respectively. Clear ΔOD signals are observed near these energies in the photoexcited sample. We focus first on the Te window in the grey box region of Figure 2a (lower panel), which is blown up for clarity in Figure 2b. At energies above 42 eV within the Te window, the XUV absorption spectrum involves overlapping transitions from both the Te $4d_{5/2}$ and the Te $4d_{3/2}$ core levels, as shown in Figure 1c. Below 42 eV (*i.e.* 38-42 eV), however, the spectrum is characterized by resonant transitions involving only the Te $4d_{5/2}$ core level, which greatly facilitates the analyses and interpretation of experimental spectra in this region. Two ΔOD spectral lineouts in this energy window are shown in Figure 2c, averaged over short (10-50 fs) and long delay times (2.5-3 ps). Within this energy range (38-42 eV), the resonant transitions correspond to promotion of an electron primarily from the Te $4d_{5/2}$ core level into the unoccupied DOS of the VB and CB.

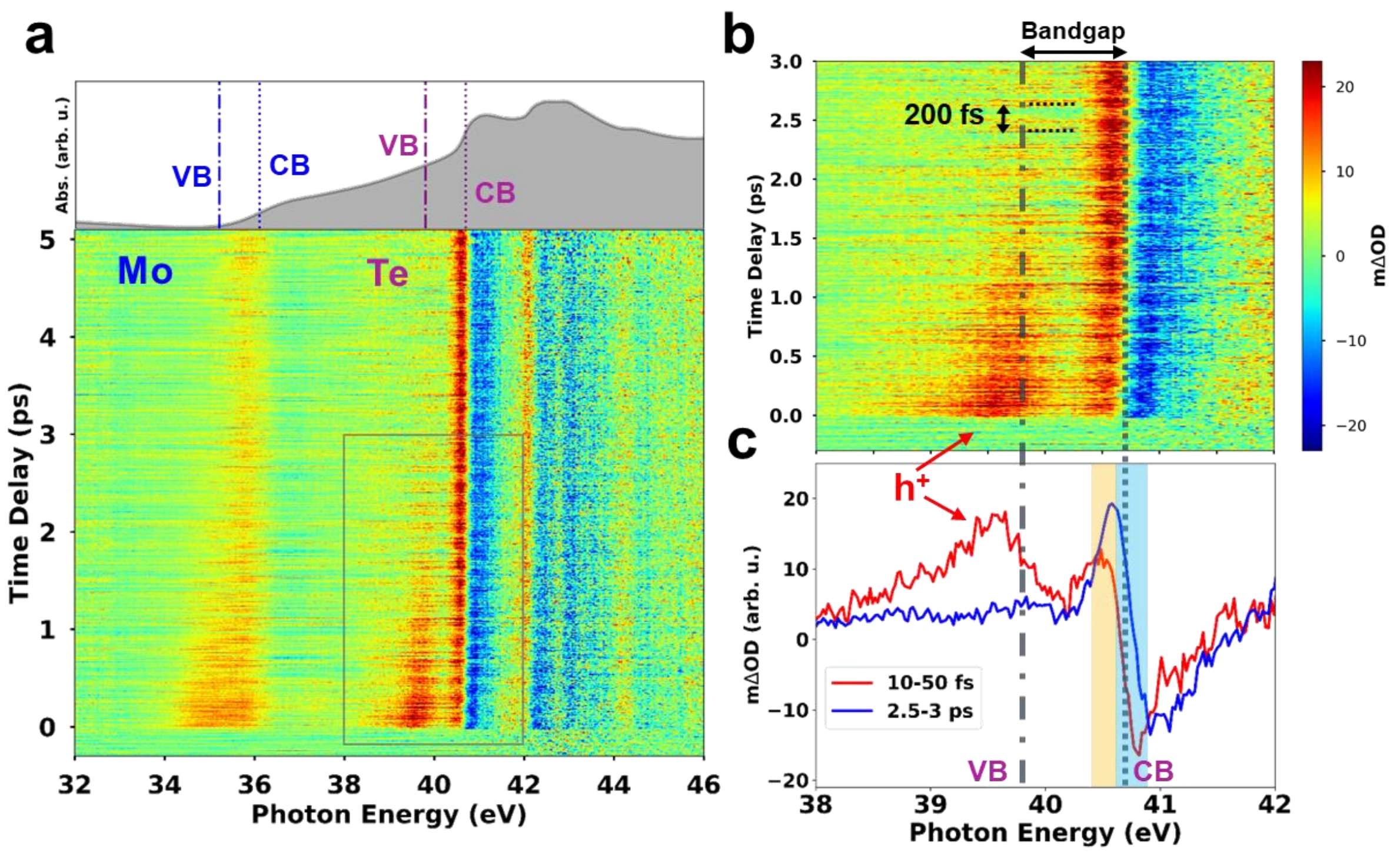

**Figure 2. (a)** XUV transient absorbance in a false-color plot *versus* time shown in the bottom panel and compared to the static absorbance spectrum in the top panel. The vertical dotted and dash-dotted lines in the upper panel are drawn at the energies corresponding to the CB minimum and VB maximum, respectively, from both the Mo $4p_{3/2}$ and Te $4d_{5/2}$ core levels. **(b)** Transient absorbance in a false color plot *versus* time, expanded from the grey box region in the lower panel of (a). The indicated 200 fs oscillations in the differential absorbance is discussed further in the final section of the paper. **(c)** Differential absorbance lineouts averaged over early time delays (10-50 fs) and long time delays (2.5-3ps). Orange and blue shaded regions indicate the energy windows used for temporal lineouts in Figure 5.

Two main features are observed in the differential spectra at early delay times (10-50 fs window): a positive peak appearing near 39.5 eV and a derivative-shaped feature centered at the onset of the ground-state $4d_{5/2}$→CB absorption edge at 40.7 eV. The positive feature centered at 39.5 eV appears 0.3 eV below the derived $4d_{5/2}$→VB maximum (shown as dash-dotted vertical line). This feature is therefore assigned to transitions from the $4d_{5/2}$ core level to 'hot' holes in the VB produced by the above-bandgap photoexcitation (schematically represented by the left panel in Figure 1d). In the later time window (2.5-3 ps), the $4d_{5/2}$→VB hole feature has disappeared and the derivative feature is characterized by a slight blue shift relative to the early delay times and a significant increase in absorption at energies between 40.6 and 40.9 eV. Several effects contribute to the derivative feature, including state-filling by the electrons in the CB and bandgap renormalization (BGR), as schematically illustrated in Figure 1d.[21,23,30] These overlapping effects, which are discussed in detail below, make the electron state-filling signal less straightforward as compared to the hole signal. In the following, we therefore concentrate first on extracting the hole distribution and population dynamics from the temporal evolution of the $4d_{5/2}$→VB signal centered near 39.5 eV.

**Hole Distribution and Population Dynamics in the Valence Band.**

In order to track the detailed dynamics of both the population and distribution of the holes in the VB, we measure the time-dependent amplitude (population) and the central energy (distribution) of the $4d_{5/2}$→VB peak. In Figure 3a, the hole population measured *via* the integrated $4d_{5/2}$→VB absorption amplitude near the VB maximum (39.7-39.9 eV) is plotted as a function of delay time. The population decay is fit to a single exponential with a time constant of $\tau_{pop}$ = 1.5±0.1 ps, which is assigned to electron-hole recombination. The fit to an exponential decay is chosen to provide a direct comparison to recent Mid-IR and THz transient absorption measurements of the carrier lifetime in 2H-phase $MoTe_2$.[16,17] In the THz study,[16] the carrier lifetime is found to be dominated by a phonon-mediated trapping/recombination mechanism, accelerated by Auger scattering, which can be reduced to a single exponential decay constant of ~2 ps. This is consistent with our result of $\tau_{pop}$ = 1.5±0.1 ps, but here we can explicitly assign this time constant to electron-hole recombination with minimal contributions from long-lived carrier traps. Any long-lived trapped holes would still be visible in the XUV transient absorption spectrum, if present in sufficient densities,[22] but no evidence for these long-lived hole states is evident after a few ps (see Supporting Information Figure S5). The population dynamics of the holes can also be independently extracted from the Mo edge by measuring the decay of the integrated Mo $4p_{3/2}$→VB signal at 35.1-35.3 eV (Supporting Information Figure S6). A time constant of $\tau_{pop}$ = 1.4±0.1 ps is extracted, which matches the measurement in the Te window within the error bars.

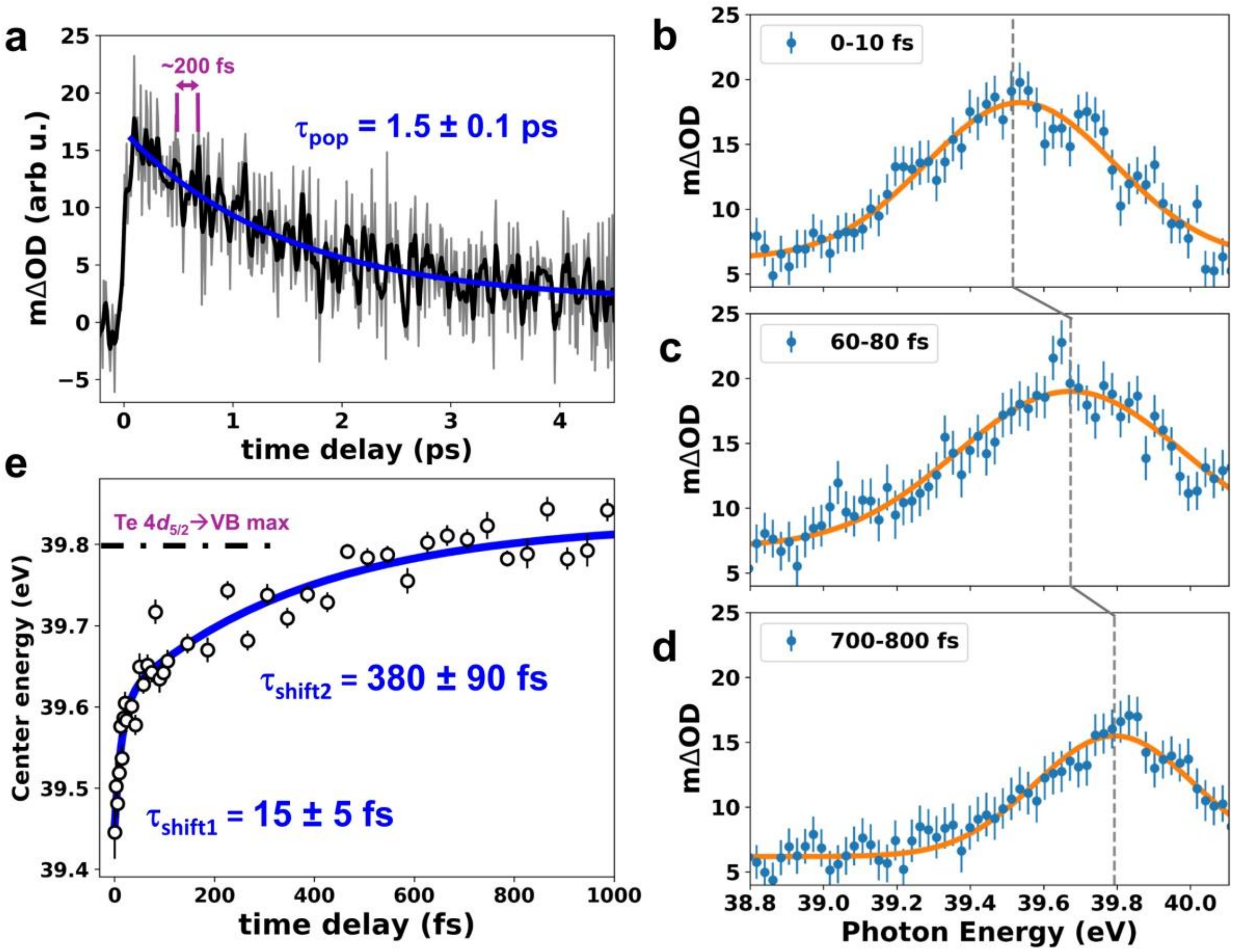

**Figure 3. (a)** Hole population dynamics measured as the integrated ΔOD amplitude at 39.7-39.9 eV *versus* time delay. The grey solid line is the experimental absorption data and the black solid line is a rolling average over 7 delay points. A single exponential decay is fit to the data (blue solid line) with a time constant of 1.5±0.1 ps **(b)-(d)** Differential absorption of the Te $4d_{5/2}$→VB hole signal at time delays of 0-10 fs, 60-80 fs, and 700-800 fs. The closed circles are the experimental absorption data and the solid orange lines are Voigt fits. The grey dashed vertical lines show the center energy of the Voigt fit in each time slice. Blue arrows show the energy shift observed between time slices. Each differential absorption spectrum is an average over 10 time points within the designated time window. At each time point, 36 ΔOD spectra are averaged and the error bars are calculated as the standard error of the mean. The error for each time point is then propagated in the average of the 10 time points for each window and plotted with the data. **(e)** Hole distribution dynamics determined by extracting the center energy of the Te $4d_{5/2}$→VB hole signal using a Voigt fit at each time delay. The open circles are the extracted center energies and the error bars represent the standard error of the center energy fitting parameter in each Voigt fit. A biexponential decay function is fit (solid blue line) to the data with time constants of 15±5 fs and 380±90 fs.

Turning now to the distribution dynamics of the holes within the VB, we first examine more closely the Te $4d_{5/2}$→VB hole feature at three representative time slices of 0-10 fs, 60-80 fs, and 700-800 fs plotted in Figure 3b-d. The spectrum of the hole feature is interpreted in accordance with the one-particle picture to directly record the hole energy distribution within the VB. In each time slice, the Te $4d_{5/2}$→VB peak is fit by a Voigt function to take into account the photoexcitation bandwidth, which produces a broad hole distribution approximated by a Gaussian function, and the convolved Lorentzian core-hole lifetime broadening. In the first time slice (0-10 fs), the $4d_{5/2}$→VB signal is centered at 39.5 eV, which is 0.3 eV below the $4d_{5/2}$→VB maximum. This peak is assigned to $4d_{5/2}$ core transitions to the nascent, non-thermal hole distribution in the VB, which is represented by the schematic illustration in the left panel of Figure 1d. In the 60-80 fs time slice in Figure 3c, the $4d_{5/2}$→VB hole feature has blue shifted from 39.5 eV to 39.65 eV. This shift of the holes to higher energies is interpreted as a combination of carrier-carrier (*i.e.* hole-hole or hole-electron) thermalization and partial hole-phonon cooling of the nascent hole distribution, which correlates to the middle and right panels of the schematic model in Figure 1d. Note that while thermalization by carrier-carrier scattering does not lead to a change in the total energy of the carrier distribution, the *peak* of the hole distribution still shifts toward the VB maximum by thermalization to a Fermi-Dirac distribution (see Supporting Information, Figure S7). A slight asymmetry is observed in the feature at 60-80 fs (Figure 3c), which could be an indication of the formation of a Fermi-Dirac distribution, but the peak is still fit well by a broad Voigt distribution. In the 700-800 fs time slice in Figure 3d, the hole feature is blue shifted further to the VB maximum at 39.8 eV and the overall amplitude of the peak is decreased. This 700-800 fs time slice is interpreted in accordance with the rightmost panel of Figure 1d. The blue shift toward the VB maximum represents near-complete cooling of the holes and the decay of the integrated area of the signal is caused by partial loss of hole population due to electron-hole recombination.

To map the full evolution of the hole distribution, the Voigt fitting procedure of the $4d_{5/2}$→VB peak is repeated for each time delay within the lifetime of the hole population and the extracted Voigt center energies are plotted as a function of delay time in Figure 3e. Although the hole distribution can also be fit by a Fermi-Dirac distribution at intermediate delay times to monitor the hole cooling process,[26] one can only define a carrier temperature in terms of a Fermi-Dirac distribution after thermalization has occurred. Since one of our primary goals is to distinguish the thermalization and cooling timescales of the holes, the Voigt fitting procedure used here allows us to trace the entire hole relaxation process using a single time-dependent parameter. The observed hole energy distribution dynamics in Figure 3e can be described by a biexponential energy shift with a fast time constant ($\tau_{shift1}$ = 15±5 fs) and a slower component ($\tau_{shift2}$ = 380 ± 90 fs). A monoexponential fit was also attempted, but did not result in a good agreement with our data (see Supporting Information Figure S7). The initial energy redistribution timescale ($\tau_{shift1}$ = 15±5 fs) from the biexponential fit is very similar to the <20 fs time constant measured by Nie *et al* in $MoS_2$ for carrier-carrier scattering using optical transient absorption spectroscopy.[37] In the experiment by Nie *et al*, the hole and electron dynamics are not distinguished and the optical spectrum is blind to the distinct energy distributions of the carriers. However, the <20 fs time constant is assigned to thermalization by carrier-carrier scattering, which is supported by *ab initio* calculations.

In our experiment, the $\tau_{shift1}$ = 15±5 fs measured in the peak of the carrier-specific hole distribution energy in 2H-$MoTe_2$ is assigned to a similar thermalization step *via* hole carrier-carrier scattering, as schematically represented in Figure 1d (left to middle panels). This assignment is consistent with a recent calculation of carrier-carrier scattering in monolayer 2H-

$MoTe_2$ at similar carrier densities (per layer), which predicts a thermalization time of sub-20 fs.[39] This is also consistent with carrier-carrier thermalization timescales measured in other semiconductors including lead iodide perovskite[36] and other layered materials including graphite.[40] Additionally, intra- or intervalley scattering processes involving carrier-phonon interactions may also contribute. An intervalley scattering time of 70 fs was measured, for example, in another TMDC semiconductor, $WSe_2$, *via* ultrafast angle-resolved photoemission spectroscopy at similar excitation fluences.[41] However, the hole distribution in 2H-$MoTe_2$ measured here does not reach the global maximum of the VB, which is localized at the Γ point (Figure 1b), until after the slower time constant of $t_{shift2}$ = 380 ± 90 fs. The hole distribution energy then remains constant near the VB maximum over the remaining hole population lifetime of $\tau_{pop}$ = 1.5±0.1 ps. The slower hole redistribution time of $t_{shift2}$ = 380 ± 90 fs is therefore assigned to hole-phonon cooling *via* intra- or intervalley scattering in the VB to the Γ point, allowing the carriers and lattice to come into thermal equilibrium before electron-hole recombination takes place.

**Spectral Decomposition of Conduction Band Transients.**

We now discuss the derivative feature in the transient spectra near the onset of the $4d_{5/2}$→CB absorption (~40.7 eV). Several effects contribute to this feature including bandgap renormalization (BGR) and state-filling by the electrons in the CB (illustrated in Figure 1d), excited-state broadening, and local modifications to the structural environment due to phonon excitations.[21,23,30] In Figure 4, the Te window ΔOD spectrum in the time slice immediately following photoexcitation (0-10 fs) is plotted along with a manual decomposition of the broadening, BGR, and state-filling contributions in this energy window (further details in Supporting Information). The decomposition is performed using a similar procedure to that used by Zürch *et al* on Germanium.[21]

The BGR and broadening contributions are calculated starting from the measured static XUV absorption spectrum by applying a linear energy shift and convolving the spectrum with a Gaussian filter, respectively. The state-filling effects are modeled by Voigt functions with a positive (negative) ΔOD sign for holes (electrons) and by taking into account the known spin-orbit splitting of the Te $4d_{5/2}$ and Te $4d_{3/2}$ core levels. The unknown parameters of the shift, broadening, and state-filling contributions are manually adjusted to find a qualitative match with the spectrum in Figure 4. The resulting decomposition shows that the Te $4d_{5/2}$ →VB hole signal is clearly isolated from the other transient effects, which makes the interpretation of this feature straightforward (as exploited in the previous section). The situation is less straightforward for the electron signal (state-filling: Te $4d_{5/2}$ →CB), which overlaps with the Te $4d_{3/2}$ →VB hole signal and the BGR and broadening contributions of the Te $4d_{5/2}$ →CB edge. Using spectral decomposition to extract the electron state-filling signal is possible at short delays when the signal is strong, as shown in Figure 4, but as the carriers recombine at later times, the decomposition becomes less reliable. In the next section, therefore, we analyze the time-dependent ΔOD observed near the Te $4d_{5/2}$ →CB edge (40.65 to 40.9 eV) in terms of both the decay in this electron state-filling contribution and the overlapping evolution of the broadening and BGR related to recombination.

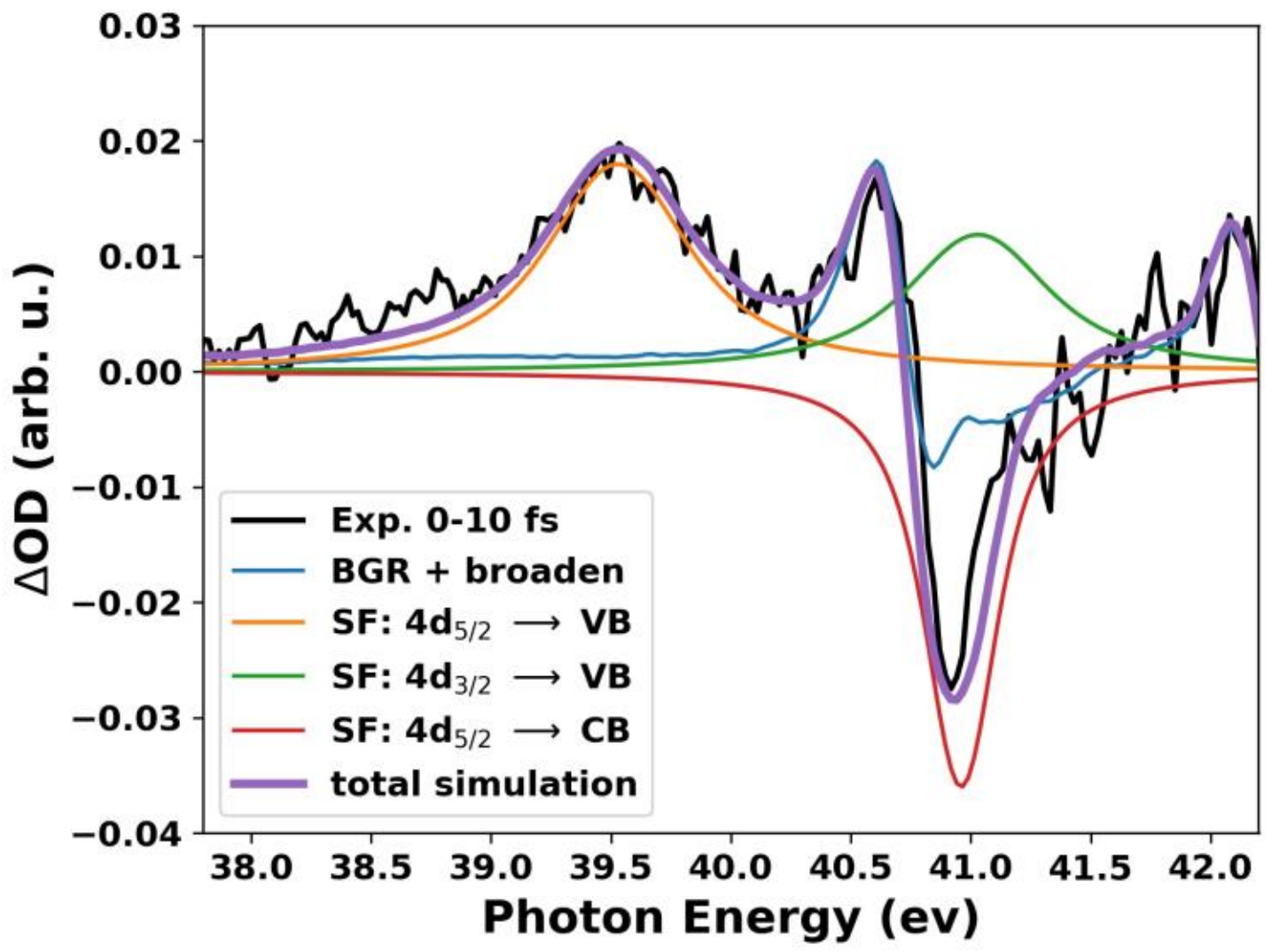

**Figure 4.**The solid black line is the experimental transient absorption data at a time delay of 0-10 fs. The colored solid lines are the manually decomposed contributions from the state-filling (SF), broadening, and bandgap renormalization (BGR), which are determined by the procedure described in the main text with further details in the Supporting Information. The total sum of the decomposed contributions is shown as the solid purple line.

**Electron Population Dynamics and Recombination-Induced Lattice Heating.**

In Figure 5a, the integrated differential absorption amplitude at energies near the onset of the $4d_{5/2}$→CB edge (40.65 to 40.9 eV, marked as blue shaded region in Figure 2c) is plotted as a function of delay time. The initial decrease in absorption at t=0 is followed by a slow increase, leading to an overall positive differential absorption after ~2 ps. The slowly evolving increase is fit to a single exponential growth and a time-constant of 1.6±0.1 ps is extracted. This time constant matches well with the decay of the hole signal shown in Figure 3a. Two overlapping effects can contribute to the increase in absorption between 40.6 to 40.9 eV, both of which are caused by electron-hole recombination. The first is the electron kinetics in the CB. As illustrated schematically in Figure 1d and demonstrated by the spectral decomposition in Figure 4, the initial state-filling effect in the CB leads to a decrease in absorption at these energies immediately following photoexcitation. When electron-hole recombination occurs, this causes the re-opening of these states with a corresponding increase in the $4d_{5/2}$→CB absorption. Second, non-radiative electron-hole recombination leads to a concomitant increase in the lattice thermal energy. Structural modifications caused by lattice heat can lead to a corresponding increase in the $4d_{5/2}$ core-level transitions at energies just below the CB minimum due to shifting and broadening effects.[23,42] This is evidenced by the increase in differential absorption at these photon energies in a continuously heated sample compared to the room temperature sample (details in Supporting Information Figure S5). Since both of these recombination-induced effects can lead to the observed increase in the XUV absorption at 40.65-40.9 eV, we assign this time-dependent increase as a signature of the electron population decay and corresponding non-radiative lattice heating.

**Coherent Lattice Displacement Dynamics:**

Superimposed on the slowly-varying ΔOD signal described in the previous section, high-frequency oscillations in ΔOD are evident near the derivative feature. In Figure 5b, the ΔOD at XUV energies where the largest-amplitude oscillations are observed (40.45 – 40.65 eV, marked as orange shaded region in Figure 2c) is plotted as a function of delay time. The inset in Figure 5b shows the high-frequency oscillations at 40.45 – 40.65 eV after subtraction of the slowly varying signal. In Figure 5c, the Fourier transform (FT) spectrum of the time-domain oscillations in the Figure 5b inset is plotted, showing the presence of two clear vibrational frequencies at 169 $cm^{-1}$ (5.1 THz) and 122 $cm^{-1}$ (3.7 THz). These frequencies match closely to the $A_{1g}$ out-of-plane and $E_{1g}$ in-plane optical phonon modes of 2H-$MoTe_2$, respectively, as observed by Raman scattering[43] (Supporting Information Figure S2). The phonon motions associated with these modes are shown schematically in Figure 5c.

From the FT spectrum in Figure 5c, the amplitude of the coherent oscillations in the XUV spectra is dominated by the $A_{1g}$ mode, which is similar to the observation of coherent phonons through transient optical transmission measurements in $MoS_2$ and $WSe_2$.[44,45] We note that the $E^1_{2g}$ mode is not observed in the FT spectrum at the characteristic frequency seen in the Raman spectrum (232 $cm^{-1}$), however, there may be evidence for a small contribution by this mode, which we describe in the Supporting Information (Figure S10). While the largest-amplitude oscillations are observed near the Te $4d_{5/2}$→CB edge (40.45 – 40.65 eV), oscillations at the dominant $A_{1g}$ frequency are also seen in other spectral regions, including near the hole signal in the Te window and near the Mo edge (see Supporting Information Figure S11). Focusing on the 40.45 – 40.65 eV region, the oscillating signal plotted in the inset of Figure 5b can be fit (blue line) using the following function:

$$\Delta OD(t) = \alpha_1 cos(2\pi t\omega_1 + \phi_1) + \alpha_2 cos(2\pi t\omega_2 + \phi_2), \quad (1)$$

where $\alpha_n$, $\omega_n$, and $\phi_n$ correspond to the amplitude, frequency, and phase of the $n^{th}$ vibrational component. The fitted frequencies are $\omega_1$ = 5.08±0.01 THz and $\omega_2$ = 3.69±0.02 THz, in agreement with what we observe in the FT spectrum in Figure 5c, and the fitted phases are $\phi_1 = 0.95\pi \pm 0.07\pi$ and $\phi_2 = 0.1\pi \pm 0.2\pi$.

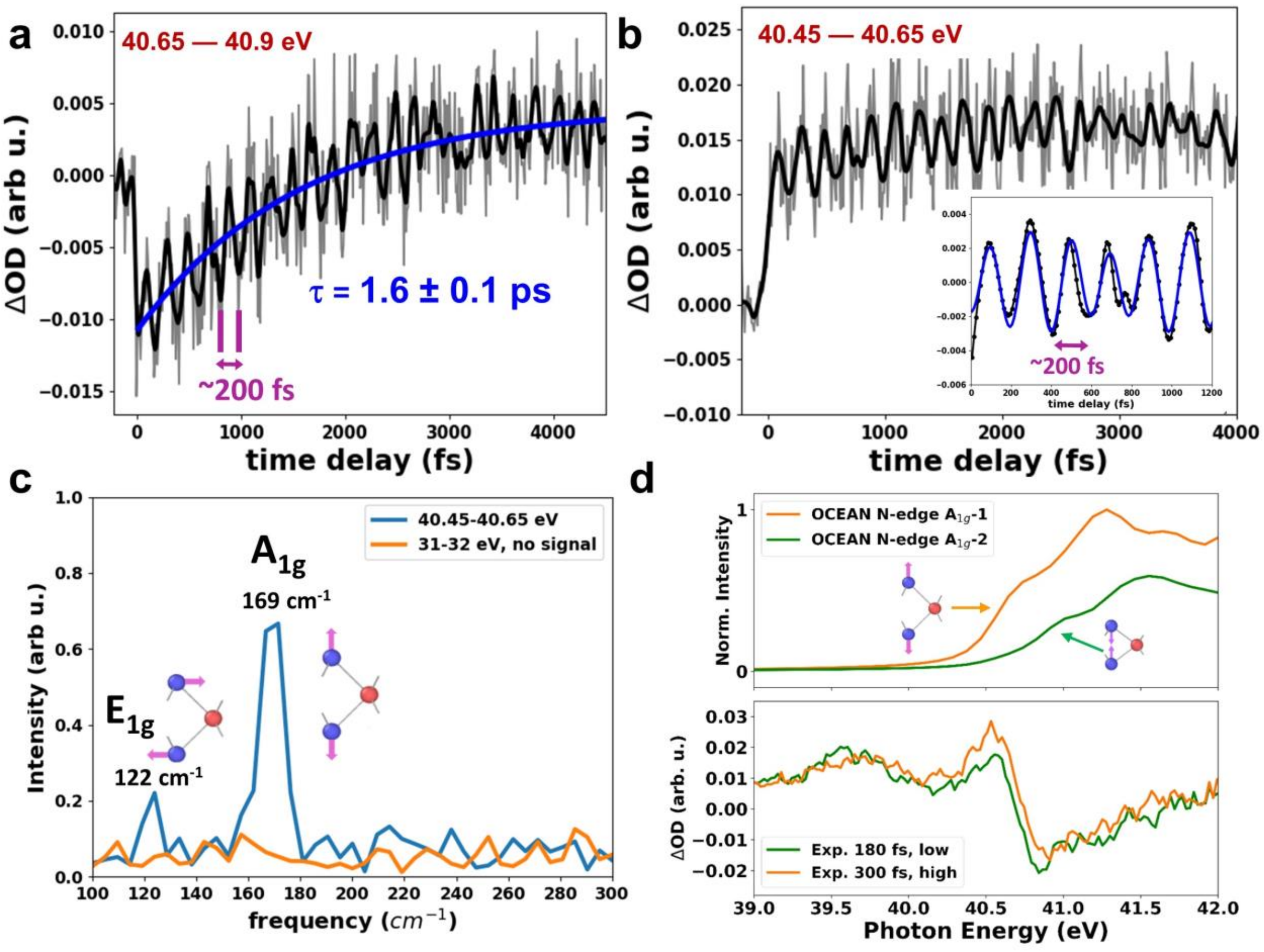

**Figure 5. (a)** The time evolution of the integrated ΔOD at photon energies just above the $4d_{5/2}$→CB edge (40.65—40.9 eV). This energy range is indicated by the blue shaded region in Figure 2c. The grey line is the experimental data and the black line is a 7-point rolling average over delay time. The blue line is a single exponential fit to the data with a time constant of 1.6±0.1 ps. **(b)** Time evolution of the ΔOD just below the onset of the Te $4d_{5/2}$→CB edge (40.45—40.65 eV). This energy range is indicated by the orange shaded region in Figure 2c. The grey solid line is the experimental absorption data and the black line shows a 7-point rolling average (smoothed) over delay time. The inset shows a cosine fit to the smoothed data as described in the main text **(c)** The Fourier transform (FT) spectrum of the time-dependent ΔOD signal in (b) is shown as a solid blue line, revealing peaks at 169 $cm^{-1}$ (5.1 THz) and 122 $cm^{-1}$ (3.7 THz). The orange line is the FT spectrum of the time-dependent ΔOD integrated over 31-32 eV (*i.e.* where no transient absorption signal is observed) to illustrate the noise floor. The inset shows a schematic of the displacive excitation mechanism, as described in the main text. **(d)** The upper panel shows the simulated displacement-induced change in the XUV spectrum calculated by OCEAN for each $A_{1g}$ distortion direction. The bottom panel shows the experimental ΔOD data at a time slices of 180 fs and 300 fs, respectively, demonstrating the uniform increase/decrease of the absorption near these energies at the two extremes of the coherent oscillation.

There are two possible mechanisms for the observed coherent phonon generation when the optical pump pulse is resonant with the electronic excitations of the material. First, the electronic excitation can lead to a direct change in the charge density of the electronic excited state and drive the nuclei to the new minimum of the potential energy surface. This mechanism is referred to as "displacive excitation of coherent phonons" (DECP).[46] Second, the optical pump pulse can couple the ground-state electronic wavefunction to the potential energy surfaces of electronic excited states through resonant impulsive stimulated Raman scattering (ISRS), resulting in a force driving coherent vibrations of the lattice.[47] As mentioned by Trovatello *et al*,[45] a consensus as to the dominant mechanism in the absorptive regime has not been established in the literature. Both mechanisms can produce displacive coherent phonons with cosine phases of 0 or $\pi$.[48,49] However, while the DECP mechanism favors only coherent phonon excitations of totally symmetric modes ($A_{1g}$),[46,50] the Raman scattering mechanism allows excitation of all Raman-active modes. Therefore, the excitation of both $A_{1g}$ and $E_{1g}$ modes (Figure 5c) indicates that resonant ISRS, also termed "transient stimulated Raman scattering" in the absorptive regime,[49] contributes to the coherent phonon excitation in 2H-$MoTe_2$. Regardless of the mechanism, the force induced by interaction with the ultrashort resonant excitation pulse in 2H-$MoTe_2$ is clearly displacive and not impulsive, as determined by the extracted phases ($\phi_1 = 0.95\pi \pm 0.07\pi$ and $\phi_2 = 0.1\pi \pm 0.2\pi$). With this information, our aim is to extract the real-space lattice displacement involved.

To understand the XUV $\Delta$OD oscillations in terms of real-space lattice displacements, we first calculate distorted 2H-$MoTe_2$ structures along the dominant $A_{1g}$ displacement coordinate. This is achieved by displacing each atom, $i$, with mass $m_i$ located at $r_i$ in the $MoTe_2$ unit cell by a distance $\Delta r_i$ along the real part of the eigenvector, $e$, corresponding to the dominant $A_{1g}$ vibration mode, $q = \Gamma$, computed as $\Delta r_i = \alpha/\sqrt{m_i}\,\mathrm{Re}[e\,\exp(q \cdot r_i)]$. The amplitude of the displacement, $\alpha$, is chosen such that the maximum real-space displacement of any atom in the simulation cell is less than 0.3 Å. Next, Bethe-Salpeter equation calculations of the $A_{1g}$-distortion-dependent Te 4*d*→CB transitions are performed using the OCEAN (Obtaining Core-level Excitations using *ab initio* methods and the NIST BSE solver) software package.[51,52] The results are plotted in Figure 5d (upper panel). The calculations reveal that the XUV absorption amplitude at the $4d_{5/2}$→CB onset reaches a maximum when the Te atoms are moved outwardly in the direction away from the plane of Mo atoms. In the opposite direction, corresponding to the Te atoms moving inwardly toward the plane of Mo atoms, the XUV absorption amplitude decreases and slightly blue shifts near the $4d_{5/2}$→ CB onset. This uniform decrease/increase over the $4d_{5/2}$→CB spectral region qualitatively matches the observed ΔOD oscillations in the experiment, as shown in the bottom panel of Figure 5d. From the phase of the $A_{1g}$ oscillation (π) extracted from the experimental data in Figure 5b, the displacement at t=0 begins with increasing absorption near the $4d_{5/2}$→CB onset. This shows that photoexcitation leads to a displacement along the $A_{1g}$ mode in the outward direction. This ultrafast lattice expansion along the specific $A_{1g}$ phonon mode in the excited state provides a key experimental benchmark for understanding how light-induced changes in electronic structure can drive non-thermal structural changes in layered materials.

**Conclusions**

In a single optical pump, XUV transient absorption probe experiment, we simultaneously uncover the detailed dynamics of intraband hole relaxation, electron-hole recombination, and excited-state coherent lattice displacement in 2H-$MoTe_2$. In contrast to optical transient absorption spectroscopy where the signals correspond to convolutions of electrons and holes at various energies within the VB and CB, here we directly map the time evolution of the hole energy distribution specifically in the VB. Taking advantage of the straightforward, one-particle interpretation of the N-edge core-level absorption, especially through the Te XUV edge, the intraband hole relaxation timescales are separated from the interband electron-hole recombination. The holes are found to redistribute within the VB on two distinct timescales of 15±5 fs and 380±90 fs. The former time constant is assigned to a thermalization process by carrier-carrier scattering, which leads to a shift in the peak of the hole energies closer toward the VB maximum. The slower hole redistribution time constant of 380±90 fs is assigned to carrier-phonon cooling by intra- or intervalley scattering to the VB maximum, which is localized at the Γ critical point. These dynamics provide key insight regarding energy loss mechanisms following above bandgap excitation of 2H-$MoTe_2$ thin films, which is important for designing next-generation photovoltaic or other optoelectronic devices using this layered semiconductor nanomaterial.

The present work furthermore reveals light-induced coherent lattice displacement dynamics along the out-of-plane $A_{1g}$ and the in-plane $E_{1g}$ phonon coordinates of 2H-$MoTe_2$. These structural dynamics show that the $A_{1g}$ mode in particular is strongly coupled to the electronic excitation. The strong electron-phonon coupling in this out-of-plane mode leads to a new equilibrium structure in the excited state reached by expansion of the Te atoms away from the Mo plane. By extracting the real-space displacement of the 2H-$MoTe_2$ lattice following optical excitation, the measurements presented here provide a benchmark for understanding how structural changes in these materials are coupled to excited-state electronic structure. Our findings and experimental approach will be of benefit to the larger community that has a growing interest in light-induced lattice displacements in layered materials for applications toward optically controlled phase transitions.[13,15]

## Methods

**Sample preparation.** The $MoTe_2$ thin film was synthesized by tellurizing a Mo film at 700°C for 2 hours in a tube furnace directly on a 30 nm thick, 3x3 $mm^2$ $Si_3N_4$ window. The Mo film was first deposited onto the $Si_3N_4$ window by sputtering previous to the reaction with Te powders. The as-synthesized $MoTe_2$ film was characterized by Raman spectroscopy and X-ray photoelectron spectroscopy (XPS). The Raman spectrum in the Supporting Information Figure S2 shows several modes including $E_{1g}$, $A_{1g}$, $E_{2g}$, indicating that the $MoTe_2$ film forms the 2H phase. Figure S3 in the Supporting Information shows the XPS spectrum of the 2H-$MoTe_2$ sample studied in this work. The relative peak spacing gives the spin-orbit splitting of the $4d_{5/2}$ and the $4d_{3/2}$ core levels of 1.5 eV. The detailed sample preparation and characterization can be found in the reference by Zhang *et al.*[3] Note that the synthesized $MoTe_2$ is a large area and continuous polycrystalline thin film.

**Experimental details.** In the XUV transient absorption experiment, laser pulses centered at 790 nm with pulse energy of 1.7 mJ and pulse duration of 30 fs from a Ti:sapphire amplifier (Femtopower Compact PRO) operating at 1 kHz repetition rate were focused into a hollow-core fiber filled with 1.5 bar of Ne to generate a supercontinuum spectrally spanning between 500 nm and 1000 nm. Dispersion compensation of the broadband pulses was achieved through broadband chirped mirrors (PC70, UltraFast Innovations) and a 2 mm thick ammonium dihydrogen phosphate crystal.[53] The laser beam was subsequently split into the probe and pump arm by a broadband beam splitter with 9:1 intensity ratio. A pair of fused silica wedges in each arm was used for fine tuning of dispersion. With dispersion scan,[54] the pulse duration of the pump and the probe were characterized to be below 5 fs (Supporting Information Figure S1a). The laser beam in the probe arm was focused into a 4 mm long cell filled with Kr to generate broadband XUV light (Supporting Information Figure S1c). The driving field in the probe arm was filtered by a 100 nm thick Al filter and the XUV light subsequently focused into the sample chamber by an Au coated toroidal mirror in a 2f-2f geometry. The pump light was optically delayed and recombined with the probe by an annular mirror with a hole at 45 degree with respect to the mirror surface. To eliminate time delay drift, a transient absorption measurement on Ar $3s^3p^6np$ autoionizing states was run after each time-delay scan on the $MoTe_2$ sample.[21]

**OCEAN and Density Functional Theory Calculations.**

Density functional theory (DFT) with the projector augmented wave (PAW) method[55] implemented in the Vienna *ab initio* Simulation Package (VASP)[56,57] was used to compute the ground-state density of states for bulk-$MoTe_2$ crystals. Exchange and correlation effects are calculated using the PBE form of GGA. Wave functions are constructed using a plane wave basis set with components up to kinetic energy of 400 eV and the reciprocal space is sampled using a $3\times3\times3$ Gamma-centered mesh with a 0.05 eV Gaussian smearing of orbital occupancies. DFT simulations of lateral interfaces were performed on a bilayer $MoTe_2$ supercell containing 216 atoms, measuring 21.09 Å × 21.09 Å × 13.97 Å along the *a*-, *b*- and *c*-directions. Calculations were performed till each self-consistency cycle is converged in energy to within $10^{-7}$ eV/atom and forces on ions are under $10^{-4}$ eV/Å.

The calculation of core-level absorption spectra at the Te $M_{4,5}$ edge was accomplished with DFT and Bethe-Salpeter equation (BSE) calculations using Quantum ESPRESSO and the OCEAN software package.[51,52,58,59] The DFT-BSE calculation was conducted using norm-conserving scalar-relativistic Perdew-Burke-Ernzerhof pseudopotentials with nonlinear core correction under generalized gradient approximation and a 6x6x1 k-point meshgrid.[60–63] In the calculation, the number of bands is set to 40 and the dielectric constant is set to 12.9. Convergence was achieved with an energy cutoff of 80 Ryd and cutoff radius of 4 Bohr.

**Associated Content**

Supporting Information. Extended experimental methods, excited carrier density, sample preparation, density of states calculations, comparison of residual heat signal at 1 kHz and 100 Hz with the transient spectrum at 4-5 ps, Mo edge hole population dynamics, thermalized carriers – shift in peak maximum, comparison of monoexponential and biexponential fits to the time-dependent energy shift of the Te $4d_{5/2}$→VB hole signal, decomposition of transient spectrum (0-10 fs), possible coherent phonon motion in the $E^1_{2g}$ mode**,** and coherent phonon motion observed in “hole” signal and in Mo window.

This manuscript has been previously submitted to a pre-print server: A. R. Attar, H. Chang, A. Britz, X. Zhang, M. Lin, A. Krishnamoorthy, T. Linker, D. Fritz, D. M. Neumark, K. Rajiv, A.

## Acknowledgements

This work was supported by the Computational Materials Sciences Program funded by the U.S. Department of Energy, Office of Science, Basic Energy Sciences, under award no. DE-SC0014607. H.-T. C. and D.M.N acknowledge support from the Air Force Office of Scientific Research (No. FA9550-14-1-0154). H.-T.C. is recently supported by the W. M. Keck Foundation grant No. 046300. S.R.L. also acknowledges support by the U.S. Department of Energy, Office of Science, Office of Basic Energy Sciences, Materials Sciences and Engineering Division, under Contract No. DEAC02-05-CH11231, within the Physical Chemistry of Inorganic Nanostructures Program (KC3103) and support by Air Force Office of Scientific Research grant FA9550-19-1-0314. The XUV transient absorption experiment is funded by the Air Force Office of Scientific Research (No. FA9550-14-1-0154) and the Army Research Office (No. W911NF-14-1- 0383). Core-level absorption simulations were conducted at the Molecular Graphics and Computation Facility, UC Berkeley, College of Chemistry, funded by the National Institute of Health (NIH S10OD023532). Density functional theory simulations were performed at the Argonne Leadership Computing Facility under the DOE INCITE and Aurora Early Science programs and at the Center for High Performance Computing of the University of Southern California. We also thank Angel Garcia Esparza and John Vinson for their help with the OCEAN calculations and Romain Géneaux for the careful reading of this manuscript.

## Author contributions

A.R.A., H.-T.C., A.B., M.-F.L., U.B., D.F., S.R.L., P.V., and A.N. conceived of the experiments. X.Z. and P.A. synthesized and characterized the experimental samples. H.-T.C., A.R.A., and A.B. performed the experimental measurements and analyzed the results. H.-T.C., A.K., T.L., R.K., A.N., and P.V. performed the *ab initio* calculations. A.R.A., H.-T.C., U.B., and S.R.L. wrote the manuscript with discussion and input from all authors.

**Competing interests:** The authors declare no competing financial interests.

For Table of Contents Only

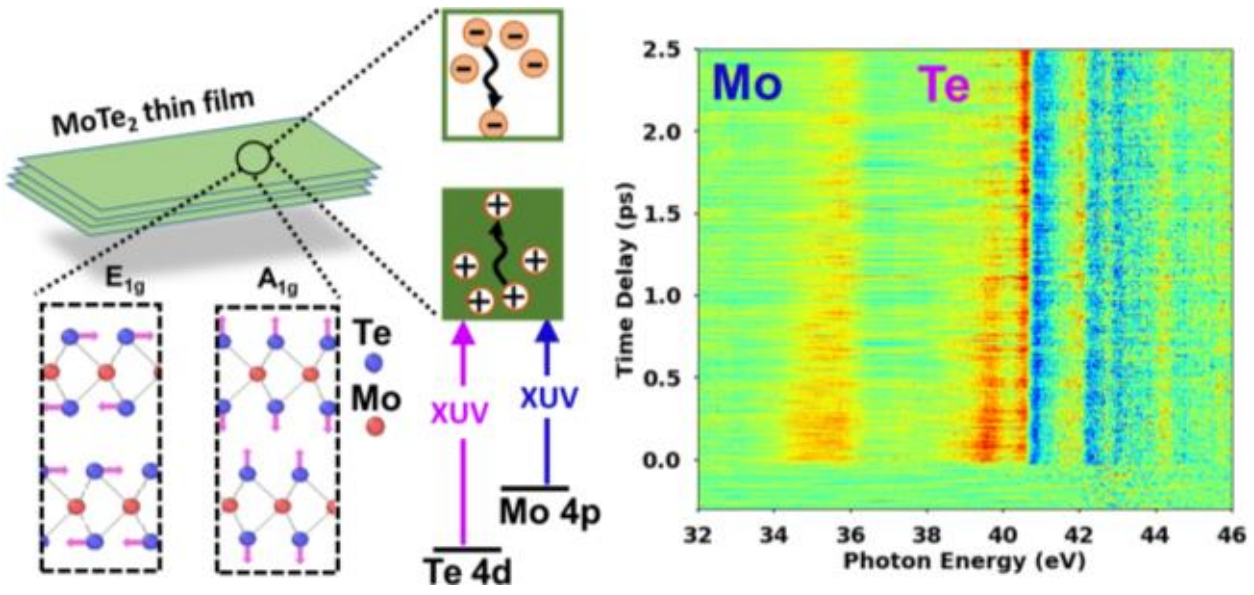

*Supporting Information for*

# Simultaneous Observation of Carrier-Specific Redistribution and Coherent Lattice Dynamics in 2H-$MoTe_2$ with Femtosecond Core-Level Spectroscopy

Andrew R. Attar[1,2,*], Hung-Tzu Chang[3,*], Alexander Britz[1,2], Xiang Zhang[4], Ming-Fu Lin[2], Aravind Krishnamoorthy[5], Thomas Linker[5], David Fritz[2], Daniel M. Neumark[3,6], Rajiv K. Kalia[5], Aiichiro Nakano[5], Pulickel Ajayan[4], Priya Vashishta[5], Uwe Bergmann[1,†], Stephen R. Leone[3,6,7†]

1. Stanford PULSE Institute, SLAC National Accelerator Laboratory, Menlo Park, CA 94025, USA.
2. Linac Coherent Light Source, SLAC National Accelerator Laboratory, Menlo Park, CA 94025, USA.
3. Department of Chemistry, University of California, Berkeley, CA 94720, USA.
4. Department of Materials Science and NanoEngineering, Rice University, Houston, TX 77005, USA.
5. Collaboratory for Advanced Computing and Simulations, University of Southern California, Los Angeles, CA 90089, USA.
6. Chemical Sciences Division, Lawrence Berkeley National Laboratory, Berkeley, CA 94720, USA.
7. Department of Physics, University of California, Berkeley, CA 94720, USA

*****These authors contributed equally to this work

† Correspondence and requests for materials should be addressed to U.B. (email: bergmann@slac.stanford.edu) or S.R.L. (email: srl@berkeley.edu)

## Extended experimental methods.

In the XUV transient absorption experiment, carrier-envelope phase (CEP) stabilized pulses centered at a wavelength of 790 nm with a duration of 30 fs and a pulse energy of 1.7 mJ are produced from a Ti:sapphire amplifier (Femtopower Compact PRO) operating at 1 kHz repetition rate. These pulses are focused into a hollow-core fiber filled with 1.5 bar of Ne to generate a visible-to-near-IR (VIS-NIR) supercontinuum spectrally spanning between 500 nm and 1000 nm. Dispersion compensation of the broadband pulses is achieved through broadband chirped mirrors (PC70, UltraFast Innovations) and a 2 mm thick ammonium dihydrogen phosphate crystal.[1] A mechanical chopper is used to reduce the repetition rate to 100 Hz to allow for thermal recovery of the sample between laser pulses (see Figure S5 for more information). The pulses are subsequently split into the probe and pump arm by a broadband beam splitter with 9:1 intensity ratio. A pair of fused silica wedges in each arm was used for fine tuning of dispersion. With dispersion scan,[2] the pulse duration of the pump and the probe were characterized to be below 5 fs (Figure S1a). In Figure S1b, the bandwidth and dispersion of the VIS-NIR pulse is shown. The pulses in the probe arm are focused into a 4 mm long cell filled with Kr to generate broadband XUV light (Figure S1c) as an isolated attosecond XUV pulse or a short attosecond pulse train. The driving field in the probe arm was filtered by a 100 nm thick Al filter and the XUV light subsequently focused into the sample chamber by an Au coated toroidal mirror in a 2f-2f geometry. The pump light was optically delayed and recombined with the probe by an annular mirror with a hole at 45 degree with respect to the mirror surface. The pump-probe time delays are scanned multiple times (36 cycles) in one experiment to improve signal-to-noise. To eliminate time delay drift over the course of the experiment (~12 hours), a transient absorption measurement on Ar 3*s*3*p*6n*p* autoionizing states was run after each time-delay cycle on the $MoTe_2$ sample (for details, see reference by Zürch *et al*).[3]

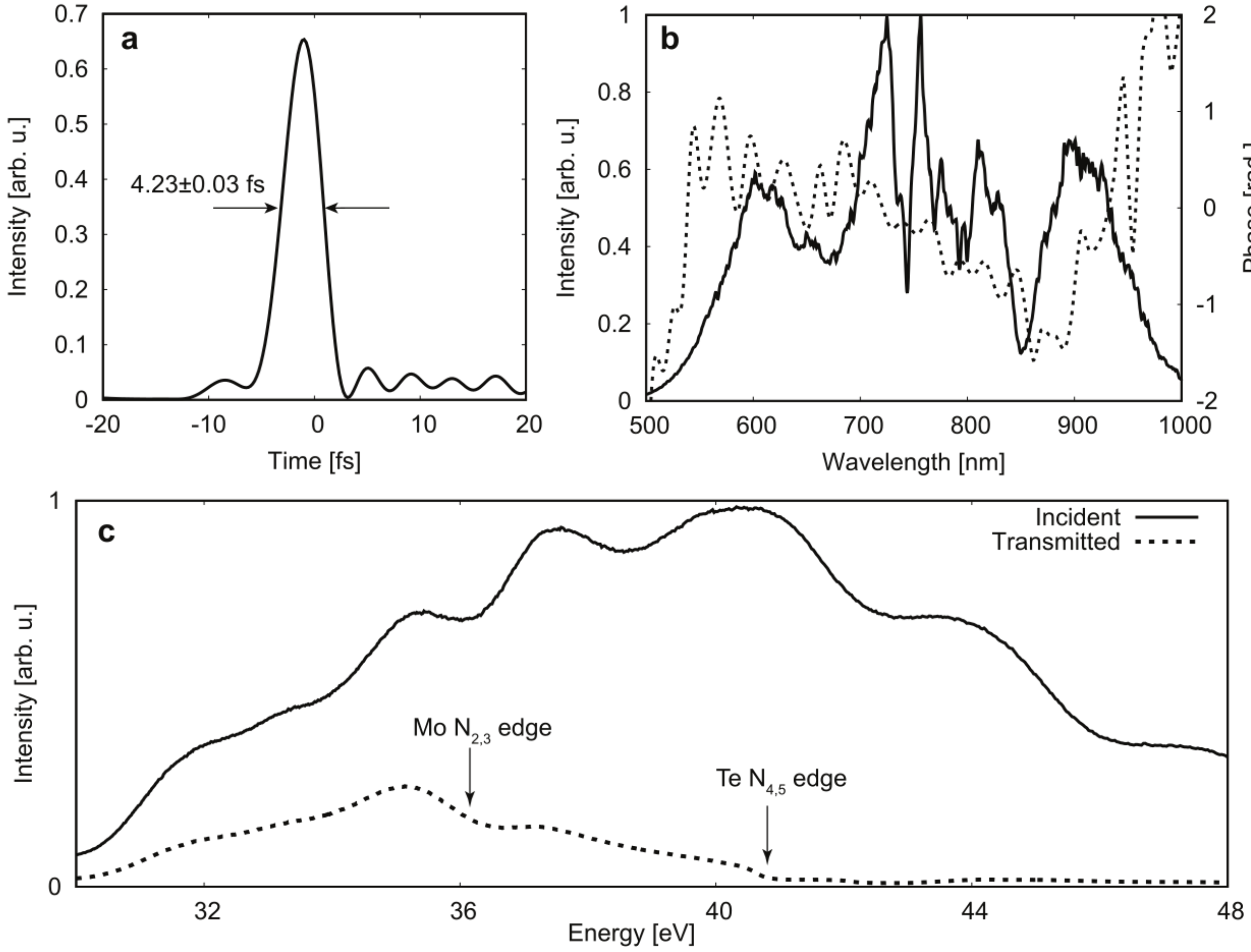

**Figure S1. Pump-probe pulse characterization. (a)** VIS-NIR pulse duration measured *via* dispersion scan method. **(b)** The spectral bandwidth and dispersion of the VIS-NIR pulses. **(c)** XUV spectrum with and without the $MoTe_2$ sample. The energies of the Mo $N_{2,3}$- and the Te $N_{4,5}$-edges are indicated.

**Excited carrier density**

The carrier density is determined by estimating the total photon absorption per pump excitation pulse, assuming each absorbed photon produces one excited electron – hole pair. In the measurements presented here, the pulse energy is 1.2 μJ and the spot size of the pump pulse at the sample is 2ω = 190 μm. We use the equations described in chapters 4(119) and 4(120) in the reference by Heavens,[4] which take into account the complex refractive indices of air, the 2H-$MoTe_2$ sample (50 nm thick), and the $Si_3N_4$ substrate (30 nm thick) to calculate the percent absorption, reflection, and transmission. This gives a total percent absorption of the incoming excitation light of 26.5% in 2H-$MoTe_2$ before taking into account the saturable absorber effect. In the next step, a correction due to the saturable absorber effect is calculated using:

$$\alpha = \frac{\alpha_0}{1 + \frac{I}{I_s}}$$

Where $\alpha_0$ and $\alpha$ are the absorption coefficients with and without the saturable absorber effect, respectively. I is the peak intensity used in the experiment and $I_s$ is the saturation peak intensity. With $I = 1700$ GW/cm$^2$ and $I_s = 217$ GW/cm$^2$,[5] we find $\alpha = 0.113$, which gives a final absorption percent of 26.5% × 0.113 = 3%. The carrier density is therefore estimated to be $1.02\times10^{20}$ cm$^{-3}$. Considering 2H-$MoTe_2$ has a unit cell volume of $1.7395416\text{x}10^{-22}$ cm$^3$, with 18 valence electrons per unit cell, the total valence electron density is $1.034755 \text{ x } 10^{23}$ cm$^{-3}$. This gives the excitation percent in the present experiment of 0.1%.

**Sample preparation**

The $MoTe_2$ thin film was synthesized by tellurizing a Mo film at 700°C for 2 hours in a tube furnace directly onto a $Si_3N_4$ membrane of 30 nm thickness and 3x3 mm lateral size. The Mo film was first deposited onto the $Si_3N_4$ window by sputtering previous to the reaction with Te powders. This results in a homogeneous polycrystalline $MoTe_2$ thin-film of ~50 nm thickness on the $Si_3N_4$ window. The synthesized samples are characterized by Raman spectroscopy and X-ray photoelectron spectroscopy (XPS). In the Raman spectrum in Figure S2, the $E_{1g}$, $A_{1g}$, and $E_{2g}$ phonon modes are observed at the characteristic frequencies of the 2H phase of $MoTe_2$ and labeled on the Figure. In Figure S3, the XPS spectrum of the 2H-$MoTe_2$ sample studied in this work is plotted. While the absolute scale is shifted slightly, the relative peak spacing gives the spin-orbit splitting of the $4d_{5/2}$ and the $4d_{3/2}$ core levels. Here we measure a splitting of 1.5 eV, which is identical to the $4d_{5/2,3/2}$ spin-orbit splitting of atomic Te.[6]

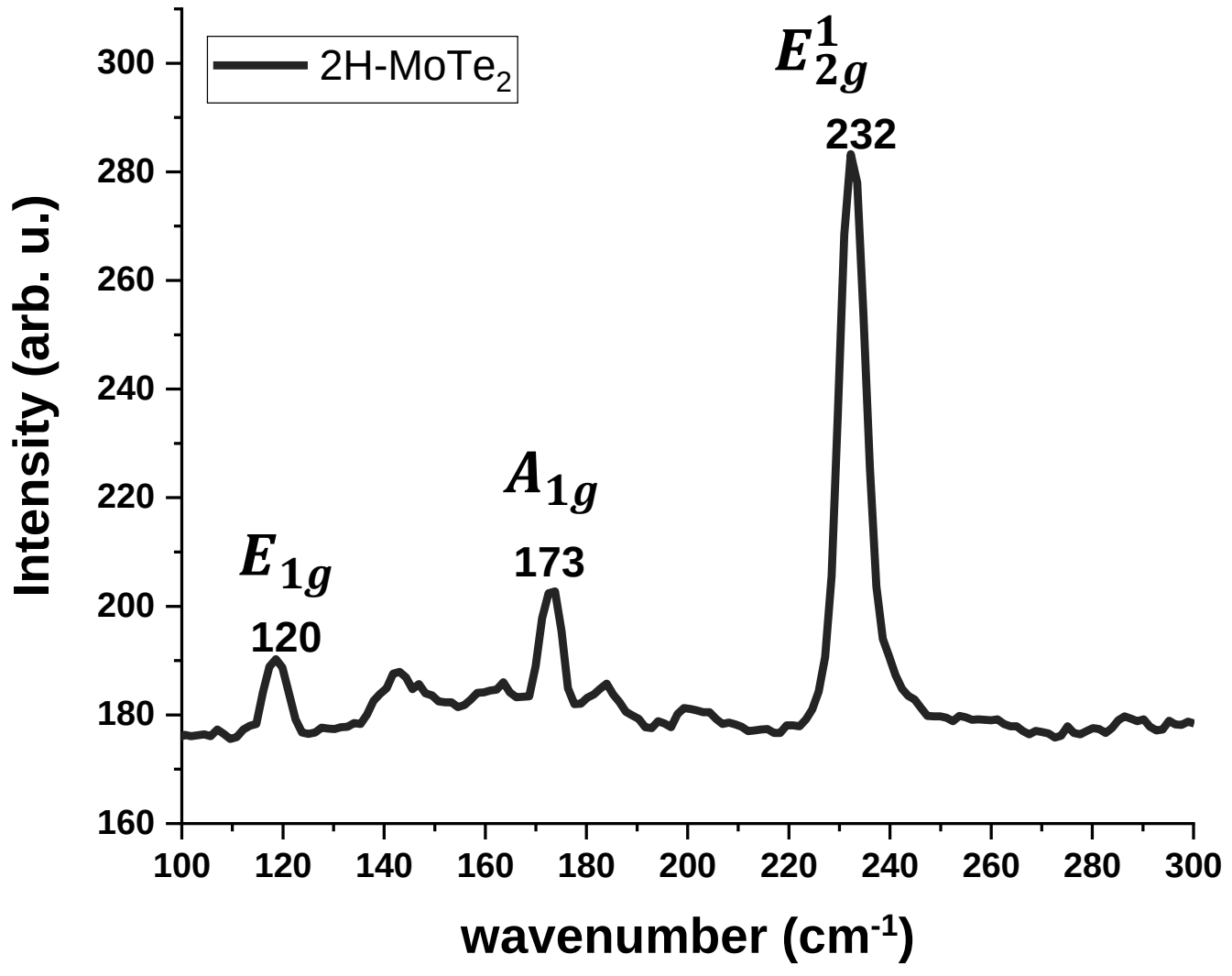

**Figure S2** Raman Spectrum of 2H-$MoTe_2$ samples

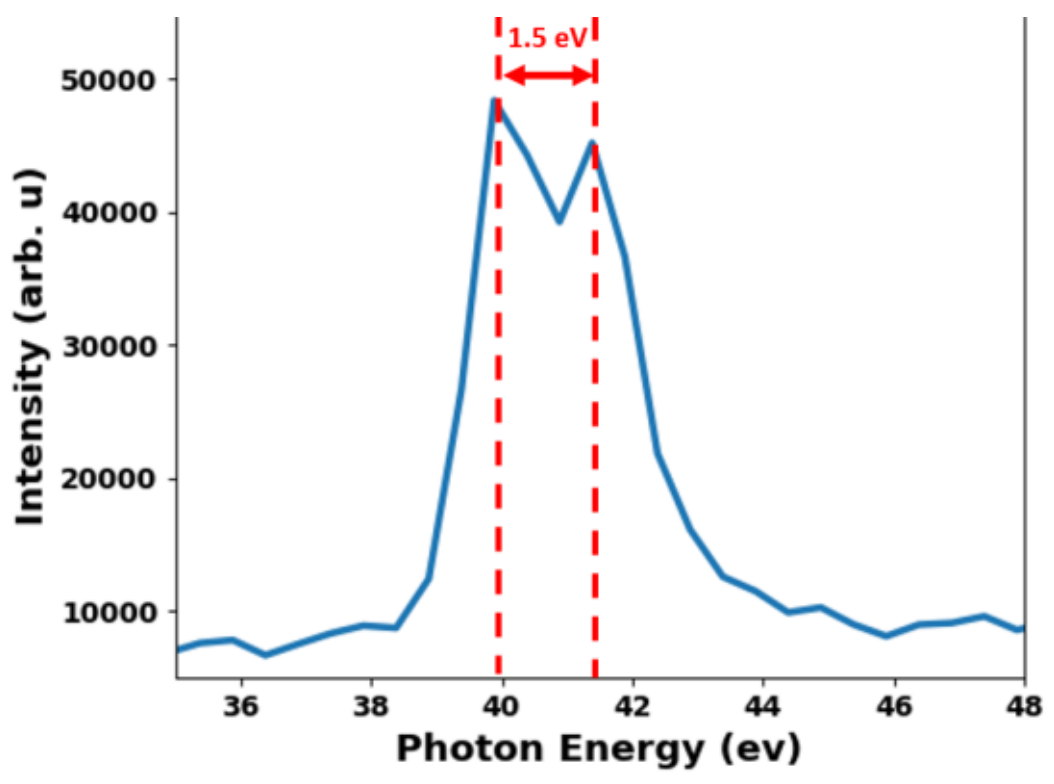

**Figure S3** XPS spectrum of 2H-$MoTe_2$ near the Te 4*d* binding energies. The binding energies of the $4d_{5/2}$ and $4d_{3/2}$ core levels are spin-orbit split by 1.5 eV.

## Density of States calculations

Density functional theory (DFT) with the projector augmented wave (PAW)[7] method implemented in the Vienna *ab initio* Simulation Package (VASP)[8,9] was used to compute the ground-state density of states for bulk-$MoTe_2$ crystals. Exchange and correlation effects are calculated using the Perdew–Burke–Ernzerhof form of the generalized gradient approximation to the exchange-correlation functional. Valence electron wave functions are constructed using a plane wave basis set containing components up to a kinetic energy of 400 eV and the reciprocal space is sampled using a 3x3x3 Gamma-centered mesh with a 0.05 eV Gaussian smearing of orbital occupancies. Figure S4 shows the calculated electronic density of states (DOS), including partial Mo and Te DOS as well as total DOS.

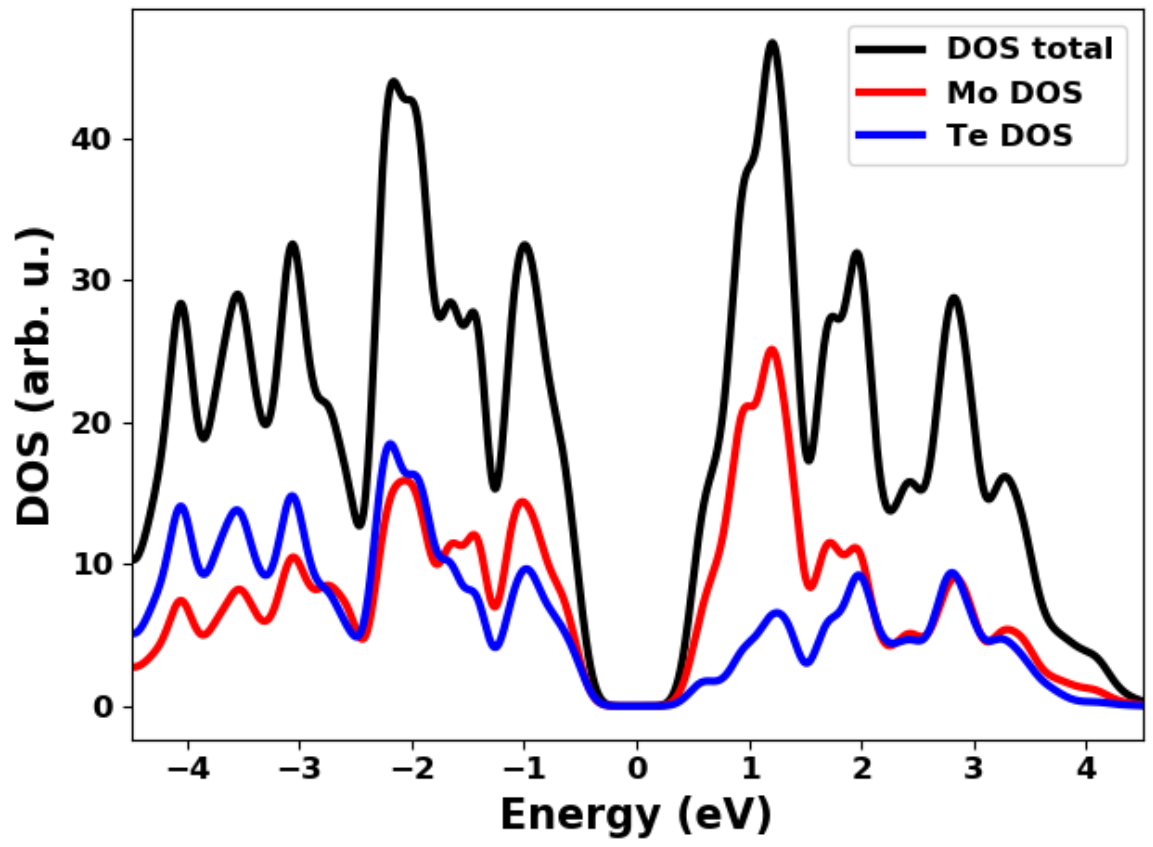

**Figure S4** Calculation showing both partial and total DOS in the VB and CB. The DOS is dominated by Mo 4*d* and Te 5*p* orbitals.

**Comparison of residual heat signal at 1 kHz and 100 Hz with the transient spectrum at 4-5 ps.**

As discussed briefly in experimental methods section above, the repetition rate of the laser is reduced to 100 Hz in the data presented in this main text in order to allow the photoexcited sample to recover close to room temperature conditions between every laser pulse (10 ms separation). Without the chopper, the 1 ms temporal spacing between the 1 kHz pulses at 1 kHz is not sufficient to allow lateral cooling of the sample to the sample holder. In this 1 kHz mode, the heat deposited from each laser pulse therefore accumulates in the sample.[3] From previous analysis of this heat deposition effect in Germanium thin films of similar thickness and thermal conductivity,[3] it was found that a thermal equilibrium is reached after only a few laser pulses. Therefore, the two datasets shown in Figure S5 labeled in the legend as “heat”, which are collected by averaging 32 cycles of 1,000 pulse integrations at negative time delays, is a measurement of the differential absorption between this thermally equilibrated sample at an elevated temperature *versus* the room temperature sample in the absence of the pump. Comparison between the negative-time-delay spectrum at 100 Hz repetition rate *versus* the 1 kHz experiment shows the significant reduction in heat accumulated in the sample due to the 10 ms delay between pulses at 100 Hz. Comparison of the purely heat-induced change in the XUV absorption at 1 kHz (shown as the red line in Figure S5) and the transient spectrum at 4-5 ps (blue line) indicates that the carrier relaxation process is completed by 4-5 ps and the remaining signal after this delay time is characterized dominantly by heat-induced effects. Therefore, any contribution of long-lived hole traps in the carrier relaxation process observed here must be minimal and is below our detection limit.

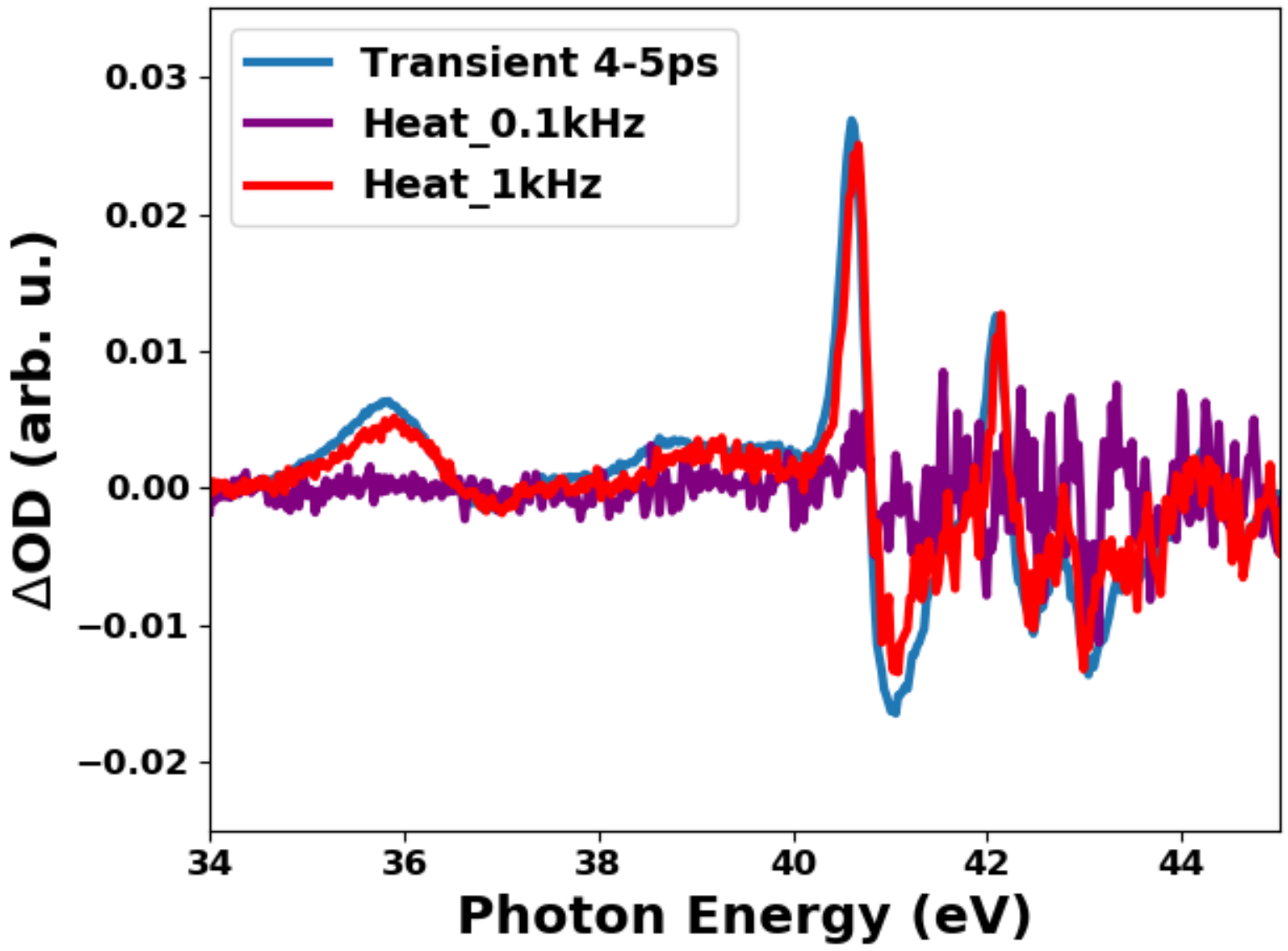

**Figure S5. Transient spectrum in the long delay time limit compared to residual heat signal.** The purple and red lines are differential absorption spectra averaged over negative time delays of -1 to -0.5 ps, taken on the same sample, but with different repetition rates of 100 Hz and 1 kHz, respectively. As discussed in the supplementary text, the negative-time-delay spectrum represents the residual changes in XUV absorption persisting from the previous pump pulse excitation (delay of 10 ms and 1 ms for 100 Hz and 1 kHz repetition rates, respectively). The residual change is due to heat that has not been sufficiently conducted away from the sample before the next pulse arrives and therefore the purple/red lines represent the difference between the XUV spectrum of this residually heated sample and the spectrum in the absence of heating (pump beam blocked). The residual heat signal is significantly reduced at 100 Hz. The blue line is the normalized differential absorption spectrum averaged over time delays of 4-5 ps from the data in the main text (100 Hz, 8 mJ/cm$^2$).

**Mo edge hole population dynamics.**

In Figure S6a, differential absorption lineouts averaged over early time delays (10-50 fs) and later time delays (2.5-3 ps) are plotted. These lineouts are identical to the lineouts in Figure 2c, except now the plot range is expanded to show both Mo and Te windows from 33-42 eV. Signatures of the photoexcited holes at early time delays are observed *via* Mo 4$p_{3/2}$→VB transitions near 35 eV and *via* Te 4$d_{5/2}$→VB transitions near 39.6 eV. Due to the sharp XUV absorption edge observed in the Te window, there is a clear separation of the Te 4$d_{5/2}$→VB absorption compared to the Te 4$d_{5/2}$→CB onset. The situation is different in the Mo window where the broad onset of the Mo N-edge leads to overlapping Mo 4$p_{3/2}$→VB and Mo 4$p_{3/2}$→CB signals. For this reason, we focus primarily on the Te 4d core-level absorption to report on the carrier and structural dynamics in the main text (*i.e.* hole relaxation extracted in Figure 3a in the main text). However, the population dynamics of the photoexcited holes can be independently extracted using both the Mo and Te elements as reporter atoms. In Figure S6b, the hole population measured *via* the integrated Mo 4$p_{3/2}$→VB signal near the Mo 4$p_{3/2}$→VB maximum (35.1-35.3 eV) is plotted as a function of delay time. A time constant of $\tau_{pop}$ = 1.4±0.1 ps is extracted, which matches the measurement in the Te window (main text Figure 3a) within the error bars.

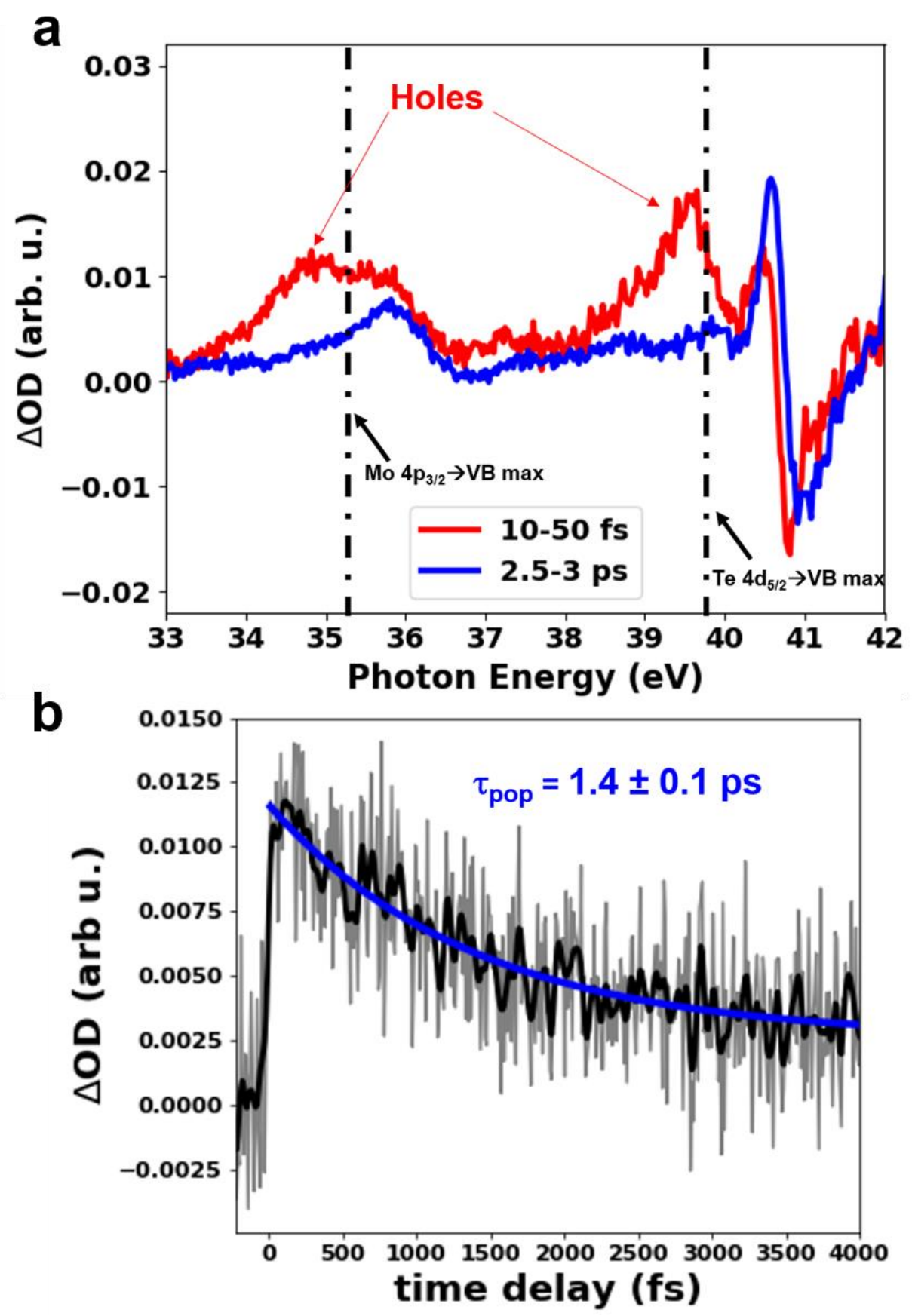

**Figure S6 Hole population dynamics measured in the Mo window. (a)** Differential absorption lineouts averaged over 10-50 fs and 2.5-3 ps. Signatures of the photoexcited holes are observed in both the Mo window and the Te window. **(b)** Hole population dynamics measured as the integrated ΔOD amplitude in the Mo window at 35.0 -35.2 eV as a function of time delay. The grey solid line is the experimental absorption data and the black solid line is a rolling average over 7 delay points. A single exponential decay is fit to the data (blue solid line) with a time constant of 1.4±0.1 ps.

**Thermalized carriers – shift in peak maximum.**

In Figure 3 of the main text, the peak of the hole distribution is observed to shift in energy toward the VB maximum with two distinct time constants, $\tau_{shift1}$ = 15±5 and $\tau_{shift2}$ = 380 ± 90 fs, which are assigned to thermalization and cooling, respectively. While it is clear that hole cooling will lead to an energy shift of the hole distribution toward the VB maximum, the energy shift effect of thermalization is less obvious. While the total energy and population of the hole distribution should not change with thermalization by carrier-carrier scattering, we illustrate using the simple model in Figure S7 that the peak position of the energy distribution does shift in energy when evolving from the non-thermal distribution to the thermalized one. The non-thermal hole distribution is modeled by a Gaussian function with a FWHM of 0.6 eV, similar to the observed Te $4d_{5/2}$→VB hole peak immediately after photoexcitation (~0-10 fs). The Gaussian distribution is centered at an energy, $E_0$, below the VB maximum, which defines the average excess energy of the hole carriers. The area of the distribution corresponds to the total number of excited holes and the integrated energy relative to the VB edge gives the total absorbed energy stored in the hole distribution. Next, we simulate a Fermi-Dirac distribution over a parabolic density of states to model the thermalized hole distribution using:

$$I(E) = \sqrt{E_{VBmax} - E} * \frac{1}{exp\left(\frac{E_f - E}{K_B T_h}\right) + 1}$$

where I is the number of carriers as a function of energy, E. The parameter $E_{VBmax}$ is the VB maximum energy, $E_f$ is the quasi-Fermi energy for the hole distribution, $K_B$ is the Boltzmann constant, and $T_h$ is the Fermi temperature of the holes. $E_f$ and $T_h$ are varied with the requirement that the total number of holes remains constant (*i.e.* no recombination) compared to the non-thermal distribution and the total integrated energy of the hole carriers also remains constant (*i.e.* carrier-carrier scattering in the absence of carrier-phonon scattering). The resulting thermalized hole distribution is plotted in Figure S7. The peak of the energy distribution is seen to shift toward the VB maximum, with a corresponding increase in the tail of the distribution toward higher energies (conserving the total energy). Therefore, tracking the peak maximum as a function of temporal delay provides sensitivity to thermalization of the carriers, even in the absence of any energy loss to the lattice.

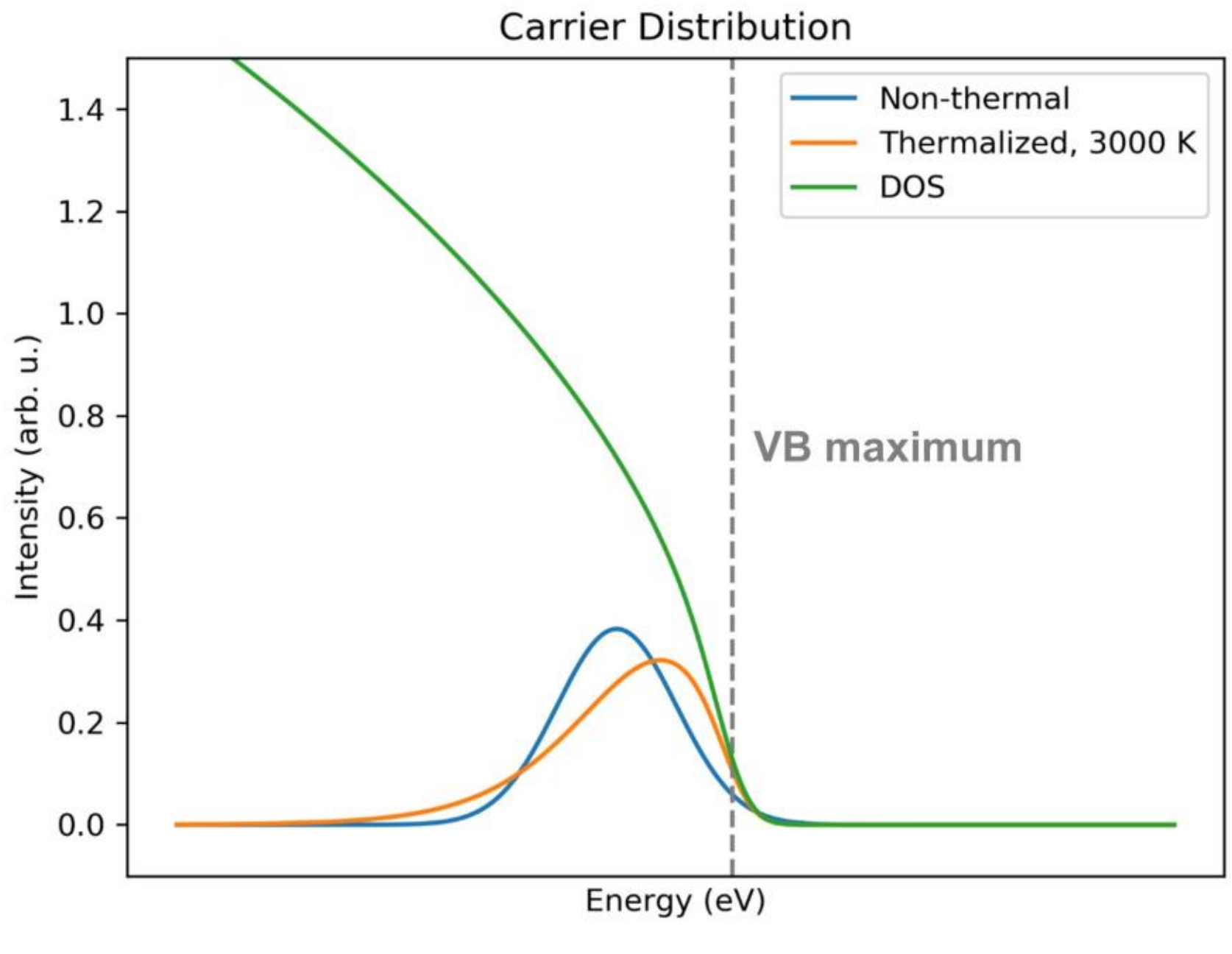

**Figure S7.** The non-thermalized distribution of carriers is represented by a Gaussian function centered at 0.5 eV below the VB maximum (blue line). The corresponding thermalized distribution (orange line) is generated as described in the text with the requirement that the total area and the total integrated energy relative to the VB maximum remains constant in comparison to the initial non-thermal distribution. The parabolic DOS is convolved with a Gaussian (s = 0.1 eV) as an approximation of the Urbach tail.

**Comparison of monoexponential and biexponential fits to the time-dependent energy shift of the Te $4d_{5/2}$→VB hole signal**

As discussed in the main text, the time-dependent energy distribution of the holes is extracted *via* a Voigt fit to the $4d_{5/2}$→VB peak at each time delay within the hole population lifetime. The extracted center energies are plotted in Figure S8 with the blue line in the left panel corresponding to a biexponential fit to the shifting energy, as shown in Figure 3e in the main text. In the right panel of Figure S8, the time evolution of the hole distribution is fit to a monoexponential function, shown as the red line. The insets of both panels show an expanded view of the early time delays up to 200 fs. The poor fit in the right panel, especially at early time delays shown in the inset, indicates that the monoexponential function is not sufficient to describe the data. A correct fit is achieved using the biexponential function in the left panel.

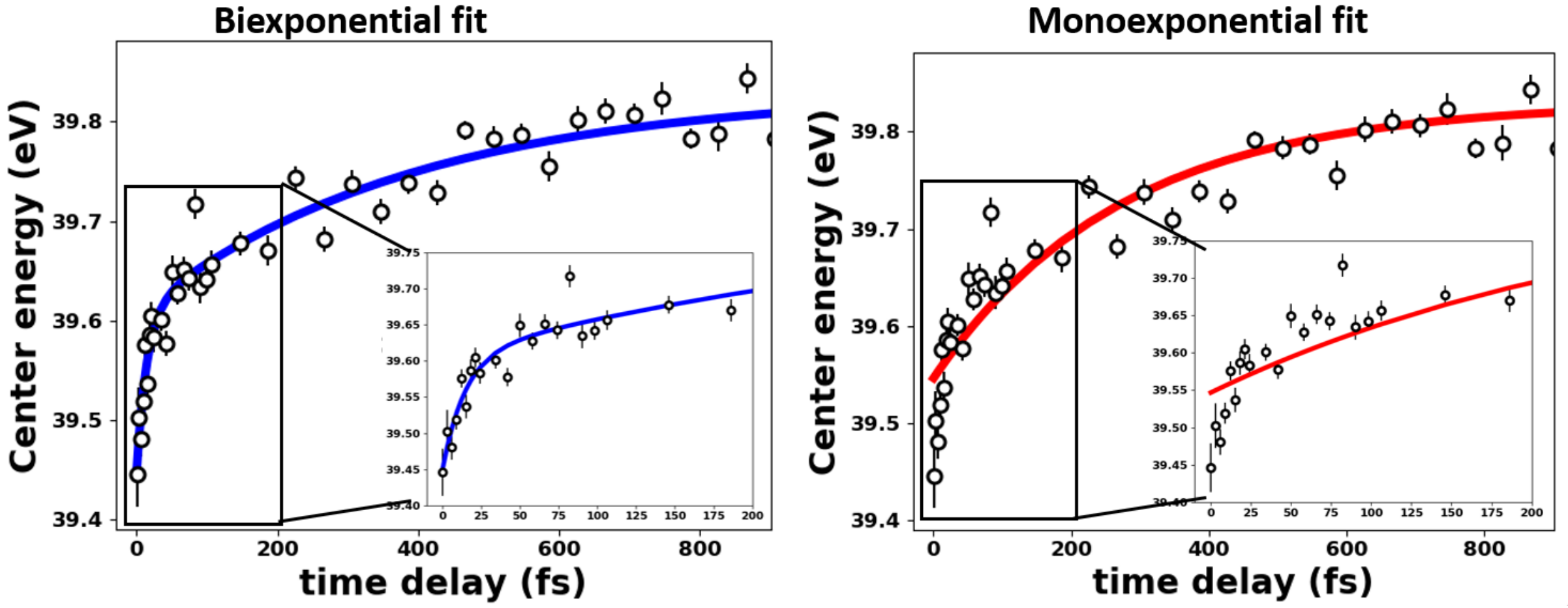

**Figure S8** Temporal evolution of the hole distribution energy and fit to a biexponential function (left) and monoexponential function (right). The left panel is the same as in Figure 3e in the main text with the addition of an inset showing an expanded view of the early time delays up to 200 fs. An expanded view of these early time delays is also shown in the right panel with the monoexponential fit. The poor fit at early time delays using the monoexponential function indicates that an additional decay component needs to be taken into account.

**Decomposition of transient spectrum (0-10 fs)**

In Figure S9, a lineout of the ΔOD spectrum near the Te $N_4$ edge averaged over delay times of 0-10 fs is plotted along with a manual decomposition of broadening, band-gap renormalization (BGR), and state-filling contributions. The decomposition is determined using the following procedure. First, state-filling (SF) by the new Te $4d_{5/2}$→VB hole transitions is manually fit to a Voigt function. The Te $4d_{3/2}$→VB hole contribution is then produced by blue-shifting the $4d_{5/2}$→VB peak by the known 1.5 eV spin-orbit splitting and by scaling down the area by the 3:2 ratio for $d_{5/2}$:$d_{3/2}$ core-hole state degeneracies. The $4d_{5/2}$→CB SF contribution (negative differential absorption) from the electrons is assumed to have the same integrated area as the $4d_{5/2}$→VB peak (*i.e.* same number of photoexcited electrons as holes) and the width of the Gaussian and center energy of the Voigt function are adjustable. The blue line shows the broadening and BGR contributions, which are manually produced by red-shifting and broadening the experimental static absorption spectrum, iterating with the adjustable parameters of the $4d_{5/2}$→CB electrons contribution, until the total simulation matches the experimental transient absorption. The resulting red shift used to simulate the BGR is 9 meV and the broadening is achieved with a Gaussian filter of σ = 0.1 eV.

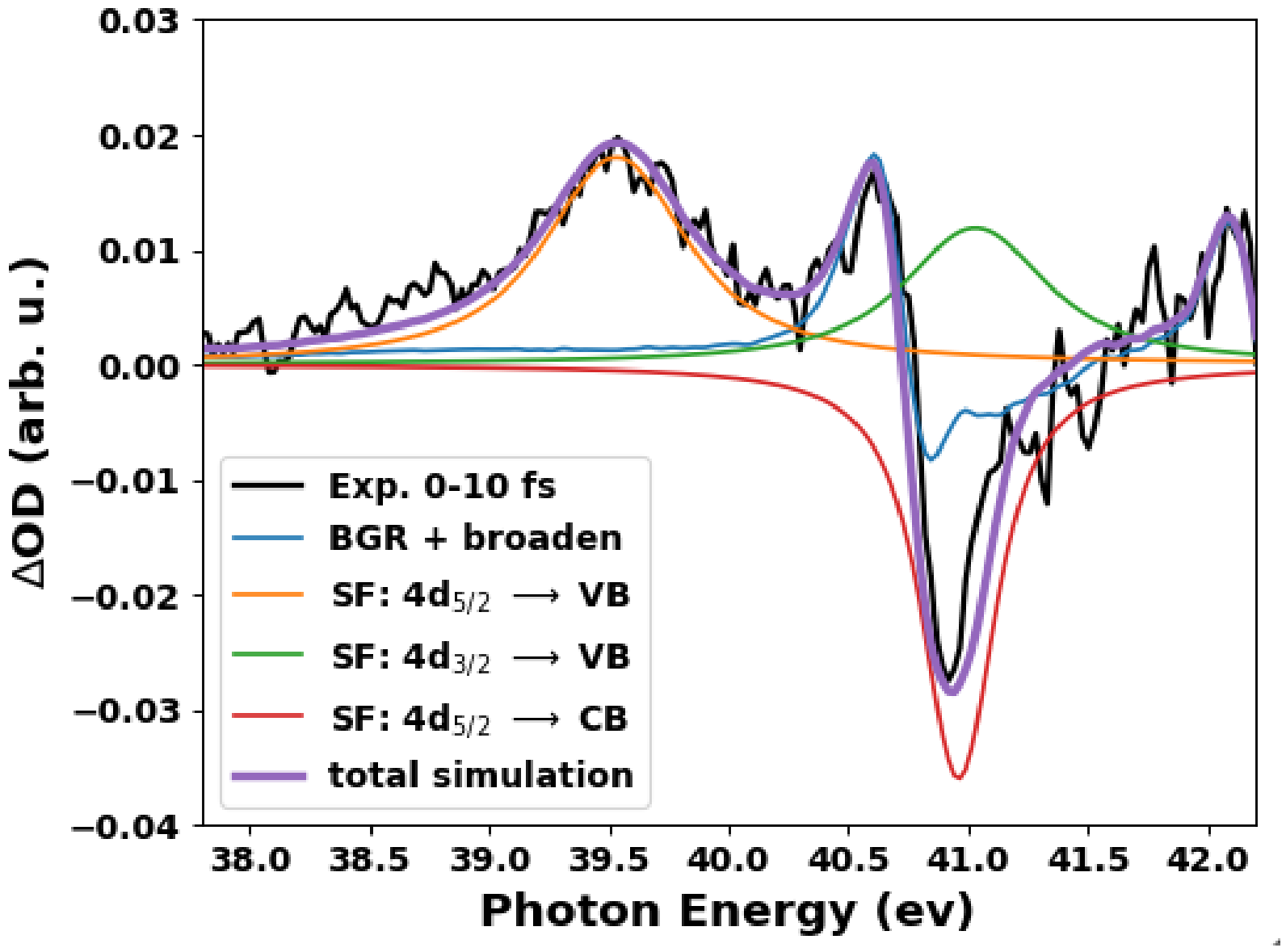

**Figure S9. Spectral Decomposition** The solid black line is the experimental transient absorption data at a time delay of 0-10 fs. The colored solid lines are the manually decomposed contributions from the state-filling (SF), broadening, and BGR. The total sum of the decomposed contributions is shown as the solid purple line.

## Possible coherent phonon motion in the $E^1_{2g}$ mode

As noted in the main text, the $E^1_{2g}$ mode is not observed at the expected frequency (~232 $cm^{-1}$) in the FT spectrum of the coherent spectral oscillations near the Te $4d_{5/2}$→CB edge. One potential reason for this is that the motion along this mode could lead to an energy shift in the XUV spectrum in the same direction for each turning point of the vibrational motion (*i.e.* at π and 2π phases). This possibility is justified by OCEAN calculations of the XUV absorption spectrum of the distorted structures along the $E^1_{2g}$ coordinate, shown in Figure S10a. At both extremes of the $E^1_{2g}$ distortion, the XUV absorption spectrum displays the same shift relative to the equilibrium structure. This may lead to an oscillation in the XUV spectrum at  double the frequency of the $E^1_{2g}$ phonon. As seen in Figure S10b, which shows the extension of the Fourier Transform (FT) spectrum in Figure 5c out to 500 $cm^{-1}$, an additional peak is observed near ~460 $cm^{-1}$, which is close to double the frequency of the $E^1_{2g}$ mode. However, this peak is near the noise level of the measurement and we feel that this does not provide conclusive evidence that the $E^1_{2g}$ mode is coherently excited. Therefore, we do not make this assignment in the present manuscript and do not include it in our model in the main text.

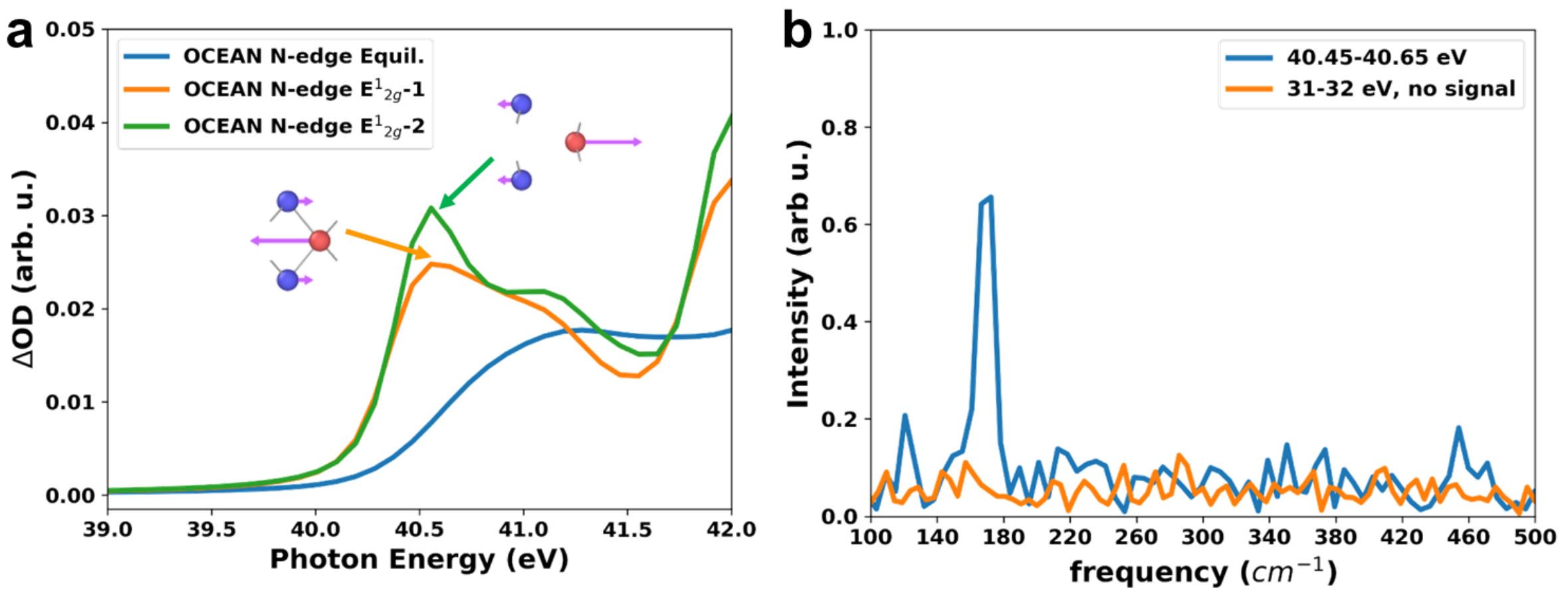

**Figure S10** (a) The simulated displacement-induced change in the XUV spectrum calculated by OCEAN for each $E^1_{2g}$ distortion direction, compared to the OCEAN-simulated XUV spectrum of the equilibrium structure. (b) The extended FT spectrum plotted in Fig. 5c of the main text, out to 500 $cm^{-1}$. A peak is observed near ~460 $cm^{-1}$, which is nearly double the expected frequency of the $E^1_{2g}$ mode (232 $cm^{-1}$).

## Coherent phonon motion observed in "hole" signal and in Mo window

In Figure S11a, the ΔOD false-color map from Figure 2a of the main text is reproduced. Thin grey lines are drawn at two energy regions, labeled (I): 35.7-36.1 eV and (II):39.7-39.9 eV. Energy (I) represents the energy region near the Mo $4p_{3/2}$→CB edge and energy (II) represents the energy region near the Te $4d_{5/2}$→VB maximum. Time-domain oscillations are observed at both energy regions (I) and (II), similar to those seen near the Te $4d_{5/2}$→CB minimum, which is described in detail in the main text. In Figure S11b-c, Fourier transform (FT) spectra of the ΔOD oscillations at energies (I) and (II) are shown, respectively. Both FT spectra show the presence of a vibrational frequency at 167 $cm^{-1}$, which matches the 169 $cm^{-1}$ $A_{1g}$ mode observed at the Te $4d_{5/2}$→CB edge in Figure 5c of the main text. This indicates that the dominant $A_{1g}$ coherent motion launched upon photoexcitation affects multiple regions of the XUV spectrum and is not specific to the Te $4d_{5/2}$→CB edge.

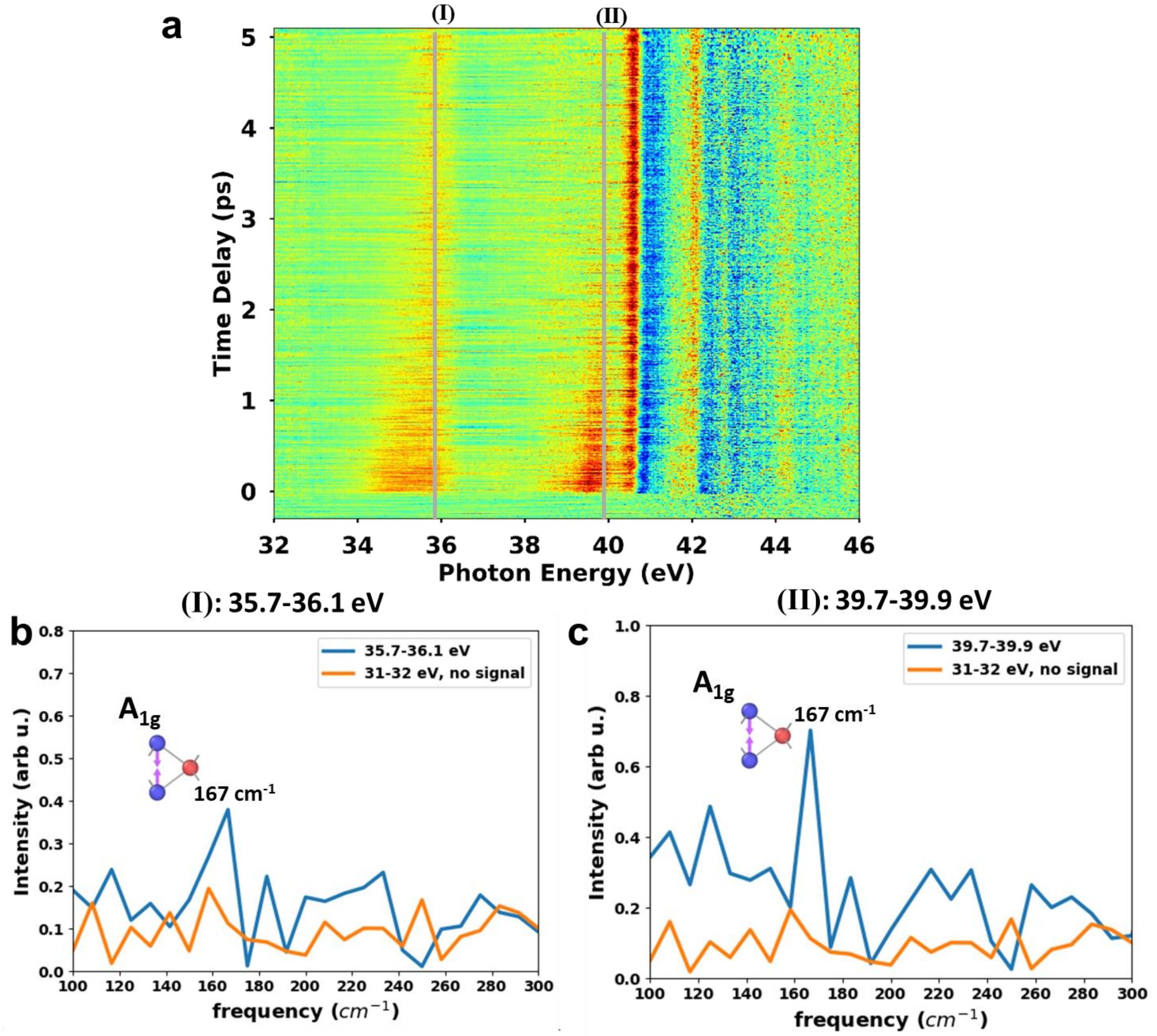

**Figure S11. Coherent phonon dynamics. (a)** XUV transient absorption false-color plot reproduced from Figure 2a in the main text. Vertical grey lines are drawn, labeled (I) and (II), which indicate the energies at which time-domain lineouts are taken. **(b),(c)** Fourier transform spectra of the time-domain lineouts taken at energy range (I):35.7-36.1 eV and (II):39.7-39.9eV. Energy (I) is taken at the onset of the Mo 4$p_{3/2}$→CB edge and energy range (II) is taken at the Te 4$d_{5/2}$→VB maximum (hole signal). Both FT spectra at these two energy regions show a peak at 167 cm$^{-1}$, consistent with the $A_{1g}$ phonon mode.